\documentclass[twocolumn]{aastex631}

\usepackage{rotating}
\usepackage{amsmath,color,graphicx,tablefootnote,enumitem}
\let\originalAA\AA
\DeclareRobustCommand{\AA}{\ifmmode\text{\originalAA}\else\originalAA\fi}
\hypersetup{pdfauthor={Zhihui Li et al.},pdftitle={Resonantly Scattered C IV Emission in Local Star-forming Galaxies: Radiative Transfer Constraints on High-ionization Gas in Reionization Analogs}}
\usepackage{graphics,subfigure,hyperref,multirow}
\usepackage[rightcaption]{sidecap}

\newcommand{\CH}[1]{\colhead{#1}}
\newcommand{\D}{$^{\dagger}$}

\newcommand{\W}{$\lambda$}

\journalinfo{}
\shorttitle{CIV Radiative Transfer}

\begin{document}
\shortauthors{Li et al.}

\title{Resonantly Scattered \ion{C}{4} Emission in Local Star-forming Galaxies: Radiative Transfer Constraints on High-ionization Gas in Reionization Analogs
\footnote{
Based on observations made with the NASA/ESA Hubble Space Telescope,
obtained from the Data Archive at the Space Telescope Science Institute, which
is operated by the Association of Universities for Research in Astronomy, Inc.,
under NASA contract NAS 5-26555.}}

\author[0000-0001-5113-7558]{Zhihui Li}
\affiliation{Center for Astrophysical Sciences, Department of Physics \& Astronomy, Johns Hopkins University, Baltimore, MD 21218, USA}

\author[0000-0002-4153-053X]{Danielle A. Berg}
\affiliation{Department of Astronomy, The University of Texas at Austin, 2515 Speedway, Stop C1400, Austin, TX 78712, USA}
\affiliation{Cosmic Frontier Center, The University of Texas at Austin, Austin, TX 78712, USA} 

\author[0000-0002-5659-4974]{Simon Gazagnes}
\affiliation{Department of Astronomy, The University of Texas at Austin, 2515 Speedway, Stop C1400, Austin, TX 78712, USA}

\author[0000-0003-1127-7497]{Timothy Heckman}
\affiliation{Center for Astrophysical Sciences, Department of Physics \& Astronomy, Johns Hopkins University, Baltimore, MD 21218, USA}

\author[0000-0003-2491-060X]{Max Gronke}
\affiliation{Centre for Astronomy of Heidelberg University, Astronomisches Rechen-Institut, M\"{o}nchhofstr. 12-14, 69120, Heidelberg, Germany}
\affiliation{Max Planck Institute for Astrophysics, Karl-Schwarzschild-Str. 1, 85741, Garching, Germany}

\author[0000-0001-9714-2758]{Dawn K. Erb}
\affiliation{Center for Gravitation, Cosmology and Astrophysics, Department of Physics, University of Wisconsin Milwaukee, 3135 N Maryland Ave., Milwaukee, WI 53211, USA}

\author[0000-0002-0302-2577]{John Chisholm}
\affiliation{Department of Astronomy, The University of Texas at Austin, 2515 Speedway, Stop C1400, Austin, TX 78712, USA}
\affiliation{Cosmic Frontier Center, The University of Texas at Austin, Austin, TX 78712, USA} 

\correspondingauthor{Zhihui Li} 
\email{zli367@jh.edu}

\begin{abstract}
Rest-frame ultraviolet emission lines are diagnostics of hard ionizing spectra and highly ionized gas in galaxies resembling the sources of cosmic reionization. Nebular \ion{C}{4} emission traces both intrinsic line production and resonant scattering through its velocity-resolved profile. Previous studies have treated \ion{C}{4} empirically, using integrated fluxes, equivalent widths, and line ratios rather than modeling the radiative transfer (RT) shaping
the emergent profiles. We present the first systematic RT modeling of resonantly scattered \ion{C}{4} emission profiles in 18 local star-forming galaxies observed with HST/COS. Using the clumpy RT framework \texttt{PEACOCK}, idealized experiments
show that profile morphology is governed by column density, gas
kinematics, intrinsic equivalent width, and aperture-dependent recovery of scattered emission. We fit continuum-normalized spectra corrected for stellar contributions with a neural-network-accelerated Bayesian pipeline, reproducing P-Cygni-like, double-peaked, double-peaked with two absorption troughs, and blue-bump plus red-peak morphologies. Emission infilling connects these profile classes and complicates kinematic inferences, while aperture losses decouple observed net equivalent widths from intrinsic line production. Combining RT-inferred intrinsic equivalent widths with rest-UV diagnostics, we find that most
galaxies are consistent with stellar photoionization, although some
may require harder spectra or more extreme ionization conditions. Comparison with CLASSY reveals stronger associations of stellar mass and star formation rate with gas kinematics than with \ion{C}{4} column density. Our results establish \ion{C}{4} as both a tracer of hard ionizing radiation and a resonant-line probe of gas structure and kinematics, with velocity-resolved profiles providing information beyond integrated line strengths in local reionization analogs.

\end{abstract}

\keywords{Galactic winds (572), Interstellar medium (847), Interstellar line emission (844), Circumgalactic medium (1879), Radiative transfer simulations (1967)}

\section{Introduction}\label{sec:intro}
Understanding the production and escape of ionizing radiation from star-forming galaxies remains a central problem in studies of galaxy evolution and cosmic reionization. Because the intergalactic medium becomes increasingly opaque to Lyman-continuum (LyC; $\lambda \lesssim 912$ \AA) photons at high redshift, direct measurements of LyC escape are not feasible for galaxies in the epoch of reionization (EoR). A major goal of rest-frame ultraviolet spectroscopy is therefore to identify indirect tracers of ionizing photon production and escape. Among the most promising diagnostics are rest-frame UV emission lines, including Ly$\alpha$ \citep[e.g.,][]{Verhamme2017,Izotov2018,Hu2023}, the resonant doublet Mg\,{\sc ii}\,$\lambda\lambda2796,2803$ \citep[e.g.,][]{Chisholm2020,Katz2022,Seive2022,Xu2022,Xu2023,Leclercq2024,Li2025} and \ion{C}{4}\,$\lambda\lambda1548,1550$ \citep[e.g.,][]{Stark2015,Berg2019,Schaerer2022,Jung2025}. The non-resonant He\,{\sc ii}\,$\lambda1640$ emission line also provides a complementary diagnostic of the production of high-energy ionizing photons \citep[e.g.,][]{Berg2019,Mondal2025}. These lines provide empirical probes of the hardness of the ionizing spectrum, the ionization state and kinematics of the gas, and possibly the low-opacity channels through which ionizing photons may escape. With the advent of JWST/NIRSpec, such diagnostics can now be studied in unprecedented detail in galaxies approaching and within the EoR.

Among UV diagnostics, \ion{C}{4} is especially valuable because the presence of nebular \ion{C}{4} emission requires a sufficiently hard ionizing spectrum to produce and maintain C$^{3+}$ ($\gtrsim48$ eV), while its collisionally excited nature makes its strength sensitive to the temperature and density of the ionized gas. It therefore probes the hardness of the ionizing spectrum, the metallicity and evolutionary state of massive stars, the ionization parameter, and possible non-stellar ionizing sources. Early detections of \ion{C}{4} in reionization-era galaxies, including the gravitationally lensed galaxy A1703-zd6 at $z \sim 7$ \citep{Stark2015}, suggested that some EoR galaxies host radiation fields harder than those typically inferred in more metal-rich low-redshift star-forming systems. More recent JWST and deep rest-frame UV observations have reinforced this picture, revealing intense \ion{C}{4} emission in young, compact, metal-poor galaxies with extreme ionization conditions \citep[e.g.,][]{Topping2025}, as well as strong N\,{\sc v} and \ion{C}{4} P-Cygni profiles and broad He\,{\sc ii} emission in UV-bright galaxies at $z \gtrsim8$ \citep{MarquesChaves2026}. These observations establish \ion{C}{4} as a powerful empirical tracer of hard radiation fields and highly ionized gas in galaxies thought to resemble the sources of reionization.

Important progress in understanding \ion{C}{4} emission in galaxies analogous to reionization-era systems has also come from studies at intermediate redshift. The VANDELS survey has identified star-forming galaxies at $z\sim3$ -- 5 with strong \ion{C}{4} emission and other properties reminiscent of reionization-era systems. \citet{Saxena2022} identified 19 strong \ion{C}{4}-emitting galaxies at $z=3.1$ -- 4.6 whose stacked spectrum shows LyC-leaker-like properties, including young, metal-poor stellar populations, high ionization parameters, strong Ly$\alpha$ emission near systemic velocity, weak low-ionization absorption, and elevated \ion{C}{4}/C\,{\sc iii}] ratios. \citet{mascia23_civvandels} further showed that \ion{C}{4} emission in 39 VANDELS galaxies at $z\sim3$ can largely be attributed to star formation rather than AGN activity, and that these systems have high ionizing photon production efficiencies and likely large escape fractions. 

At $z\sim0.3$ -- 0.4, where escaping LyC photons can be observed directly, compact star-forming galaxies and known LyC emitters provide an important calibration sample. \citet{Schaerer2022} detected \ion{C}{4} emission in six out of eight low-redshift LyC-emitting galaxies, including all systems with $f_{\rm esc}>0.1$, and found tentative evidence that \ion{C}{4}/C\,{\sc iii}] increases with LyC escape fraction. They proposed \ion{C}{4}/C\,{\sc iii}] $\gtrsim0.75$ as an empirical criterion for identifying strong LyC leakers. Similarly, \citet{Jung2025} found intense \ion{C}{4} emission with rest-frame equivalent widths exceeding 10\,\AA\ in five compact galaxies with extremely high $O_{32}$ ratios ($O_{32}>20$)\footnote{$O_{32}\equiv[\mathrm{O\,III}]\,\lambda5007/[\mathrm{O\,II}]\,\lambda\lambda3727,3729$.}. Like $O_{32}$, the \ion{C}{4}/C\,{\sc iii}] ratio is sensitive to highly ionized gas and hard radiation fields, while the resonant nature of \ion{C}{4} provides additional sensitivity to physical conditions that regulate photon escape. Together, these low-redshift studies show that strong \ion{C}{4} emission preferentially arises in young, compact, metal-poor, highly ionized systems, making local \ion{C}{4} emitters valuable analogs of galaxies that may have contributed to reionization. 

Nevertheless, the connection between strong \ion{C}{4} emission and LyC escape is not necessarily unique and may in some cases be weak or absent. Strong \ion{C}{4} emission requires both efficient production of C$^{3+}$ and favorable conditions for generating and escaping \ion{C}{4} photons, but these conditions do not map directly onto the escape of lower-energy LyC photons that are most sensitive to neutral hydrogen. The observed \ion{C}{4} strength depends on the intrinsic ionizing spectrum, gas-phase metallicity, gas temperature and density, gas geometry and kinematics, dust attenuation, and resonant radiative transfer (RT) through C$^{3+}$-bearing gas. The escape of \ion{C}{4} photons therefore indicates some transparency in highly ionized gas, but does not uniquely constrain LyC escape; neutral hydrogen can still absorb lower-energy LyC photons even when \ion{C}{4} escapes efficiently.

Disentangling the roles of intrinsic \ion{C}{4} production and RT through the high-ionization gas requires extracting the full physical information encoded in the velocity-resolved \ion{C}{4} profile. A complete characterization of LyC escape would additionally require constraints on the lower-ionization and neutral gas. Most existing studies, however, have focused on integrated \ion{C}{4} fluxes, equivalent widths, and line ratios, which provide useful empirical diagnostics of the ionizing source and nebular conditions \citep[e.g.,][]{Stark2015,Berg2019,Saxena2022,Schaerer2022,mascia23_civvandels,Jung2025}, but do not fully exploit the resonant nature of the \ion{C}{4} doublet. \citet{Berg2019} first demonstrated empirically that nebular \ion{C}{4} can exhibit resonant transfer signatures analogous to those of Ly$\alpha$, and suggested that the escape channels probed by \ion{C}{4} may differ from those traced by neutral hydrogen. Like Ly$\alpha$ and Mg\,{\sc ii}, \ion{C}{4} photons can scatter through the surrounding gas before escaping \citep[e.g.,][]{Verhamme2006,Dijkstra2014,Gronke2015,Chang2024,Li2025}. The emergent profile is therefore shaped by both intrinsic \ion{C}{4} production and resonant transfer through C$^{3+}$-bearing gas. Features such as the peak separation, red-to-blue peak ratio, and absorption trough velocity may thus encode information about the column density, covering fraction, kinematics, and geometry of the high-ionization gas through which the \ion{C}{4} photons propagate.

Interpreting the physical information encoded in the velocity-resolved \ion{C}{4} profile requires comprehensive RT modeling. Extensive Ly$\alpha$ RT studies have shown that resonant line profiles can constrain H\,{\sc i} column density, gas kinematics, and geometry \citep[e.g.,][]{Verhamme2006,Dijkstra2014,Gronke2015,Verhamme2015,LiGronke2022,Erb2023}. By contrast, \ion{C}{4} RT modeling in star-forming galaxies remains largely unexplored. This gap is particularly important because current constraints on ionizing-photon escape rely mainly on \ion{H}{1}-sensitive diagnostics such as Ly$\alpha$ or direct LyC measurements near the Lyman limit, leaving its energy dependence poorly constrained. As a tracer of C$^{3+}$-bearing gas, \ion{C}{4} provides access to the highly ionized material associated with hard radiation fields and galactic outflows. Modeling its resonant transfer can therefore constrain the relative roles of intrinsic production and scattering in shaping the observed profile, while extending studies of photon escape beyond neutral hydrogen tracers alone.

In this work, we present the first systematic RT modeling of resonantly scattered \ion{C}{4} emission-line profiles in a sample of local star-forming galaxies. We analyze 18 \ion{C}{4}-emitting galaxies observed with HST/COS, whose rest-frame UV spectra have sufficient resolution and signal-to-noise to resolve detailed \ion{C}{4}\,$\lambda\lambda1548,1550$ profile structure. By modeling resonant scattering in the C$^{3+}$-bearing medium, we investigate how the observed \ion{C}{4} profiles depend on both the intrinsic properties of the ionizing sources, such as the intrinsic \ion{C}{4} equivalent width, and the properties of the surrounding gas, including column density, velocity field, and aperture effects. This approach allows us to assess what velocity-resolved \ion{C}{4} profiles reveal about high-ionization gas, feedback-driven gas motions, and the transmission of high-energy radiation through compact star-forming galaxies.

The paper is organized as follows. In Section~\ref{sec:sample}, we describe the \ion{C}{4}-emitting galaxy sample and the HST/COS observations. In Section~\ref{sec:rt_model}, we introduce the clumpy \ion{C}{4} RT framework, including the model geometry, gas kinematics, photon sources, and treatment of aperture-dependent scattered emission. In Section~\ref{sec:rt_experiments}, we use idealized RT experiments to identify the physical origin of the main \ion{C}{4} profile morphologies. In Section~\ref{sec:fitting_CIV_data}, we fit the observed \ion{C}{4} profiles with a neural-network-accelerated Bayesian inference pipeline and interpret the resulting morphological classes. In Section~\ref{sec:uv_bpt}, we combine the RT-inferred intrinsic \ion{C}{4} equivalent widths with rest-UV line-ratio diagnostics to assess the nature of the ionizing source. In Section~\ref{sec:discussion}, we discuss the broader implications of our results in the context of the CLASSY sample and outline the main limitations and future prospects. We summarize our conclusions in Section~\ref{sec:conclusions}.

\section{Galaxy Sample, UV Observations, and Spectral Reduction}
\label{sec:sample}
Our sample consists of 18 nearby star-forming galaxies with significant ($>5\sigma$) detections of \ion{C}{4}\,$\lambda\lambda1548,1550$ emission in their rest-frame ultraviolet spectra. These \ion{C}{4} emitters are predominantly low-mass dwarf galaxies, with stellar masses $M_\star\lesssim10^8\,M_\odot$, high specific star-formation rates of order $\rm sSFR\gtrsim10^{-8}\,{\rm yr^{-1}}$, and low gas-phase metallicities, typically $Z\lesssim0.1\,Z_\odot$. Their compact sizes and highly ionized conditions make them particularly useful nearby analogs of the young galaxies in which strong rest-frame UV emission is increasingly observed during the epoch of reionization. The basic galaxy properties are summarized in Table~\ref{tbl:sample}. Here we describe the HST/COS spectroscopy used in this work, together with the stellar-continuum modeling and normalization that are essential for the subsequent \ion{C}{4} RT analysis.

\begin{deluxetable*}{rcccccCCc}[t]
\tablewidth{0pt}
\tablecaption{Resonant \ion{C}{4} Emission Sample}
\tablehead{
\CH{1}       & \CH{2 } 	        & \CH{3}   & \CH{4 }         & \CH{5}   & \CH{6} 		       & 
\CH{7}	 	 & \CH{8}           & \CH{9} \\ \hline
\CH{Target}  & \CH{R.A., Decl.} & \CH{$z$} & \CH{$f_{1500}$} & \CH{12+} & \CH{log $M_{\star}$} & 
\CH{log SFR} & \CH{log sSFR}    & \CH{References} \\ [-2ex] 
\CH{ }       & \CH{(J2000)}     & \CH{}	   & \CH{}	         & \CH{log(O/H)} & \CH{(M$_\odot$)} & \CH{(M$_\odot$/yr)}	& \CH{(yr$^{-1}$)} & \CH{(Cols 5,6,7)} }
\startdata
 1. J0337$-$0502 & 03:37:44.06, $-$05:02:40.19 & 0.01352 & 7.9  & 7.46 & 7.06 & -0.32 & -7.38 & 1, 1, 1 \\
 2. J0825$+$3532 & 08:25:55.52, $+$35:32:31.90 & 0.00240 & 1.4  & 7.37 & 6.04 & -1.98 & -8.02 & 2, 3, 3 \\
 3. J0842$+$1033 & 08:42:36.48, $+$10:33:14.04 & 0.01032 & 0.6  & 7.61 & 7.01 & -1.19 & -8.20 & 4, 3, 3 \\
 4. J0934$+$5514 & 09:34:20.02, $+$55:14:28.10 & 0.00250 & 15.0 & 6.98 & 6.27 & -1.52 & -7.79 & 1, 1, 1 \\
 5. J0947$+$4138 & 09:47:18.24, $+$41:38:16.44 & 0.00464 & 0.5  & 7.73 & 6.44 & -1.79 & -8.23 & 4, 3, 3 \\ 
 6. J0954$+$0952 & 09:54:30.48, $+$09:52:12.11 & 0.00498 & 0.6  & 7.70 & 6.53 & -1.61 & -8.14 & 4, 3, 3 \\ 
 7. J1044$+$0353 & 10:44:57.79, $+$03:53:13.10 & 0.01287 & 1.7  & 7.45 & 6.80 & -0.59 & -7.39 & 1, 1, 1 \\
 8. J1131$+$5703 & 11:31:16.32, $+$57:03:58.68 & 0.00556 & 0.7  & 7.65 & 6.51 & -1.69 & -8.20 & 4, 3, 3 \\
 9. J1201$+$0211 & 12:01:22.31, $+$02:11:08.33 & 0.00325 & 0.8  & 7.45 & 6.08 & -2.30 & -8.38 & 2, 3, 3 \\
10. J1202$+$5416 & 12:02:02.49, $+$54:15:51.05 & 0.01020 & 1.1  & 7.63 & 5.22 & -1.10 & -6.32 & 4, 5, 6 \\
11. J1214$+$5345 & 12:14:02.40, $+$53:45:17.28 & 0.00297 & 1.3  & 7.67 & 6.02 & -1.94 & -7.96 & 4, 3, 3 \\ 
12. J1323$-$0132 & 13:23:47.52, $-$01:32:51.94 & 0.02246 & 1.3  & 7.71 & 6.31 & -0.72 & -7.03 & 1, 1, 1 \\
13. J1325$-$1136 & 13:25:49.00, $-$11:36:37.94 & 0.00425 & 1.6  & 7.74 & 7.00 & -1.35 & -8.35 & 7, 8, 8 \\ 
14. J1331$+$4151 & 13:31:26.88, $+$41:51:48.24 & 0.01165 & 2.0  & 7.69 & 7.16 & -0.88 & -8.04 & 4, 3, 3 \\
15. J1418$+$2102 & 14:18:51.12, $+$21:02:39.84 & 0.00855 & 1.1  & 7.75 & 6.22 & -1.13 & -7.35 & 1, 1, 1 \\
16. J1509$+$3731 & 15:09:34.08, $+$37:31:46.20 & 0.03265 & 1.6  & 7.71 & 7.78 & +0.33 & -7.45 & 4, 3, 3 \\ 
17. J1712$+$3216 & 17:12:36.72, $+$32:16:33.60 & 0.01195 & 0.7  & 7.70 & 7.05 & -1.06 & -8.11 & 4, 3, 3 \\
18. J2238$+$1400 & 22:38:31.11, $+$14:00:28.29 & 0.02060 & 0.8  & 7.59 & 6.72 & -0.77 & -7.49 & 4, 3, 3    
\enddata 
\tablecomments{Sample of nearby galaxies with significant \ion{C}{4} emission detections, ordered by increasing R.A. All objects are bright, compact dwarf galaxies with relatively low metallicities, as measured from their ground-based optical spectra. The first three columns give the target name used in this work, sky coordinates, and redshift. Column 4 gives the FUV continuum flux density at 1500 \AA, in units of $10^{-15}$ erg s$^{-1}$ cm$^{-2}$ \AA$^{-1}$. Columns 5$-$8 list the gas-phase oxygen abundance, $12+\log(\mathrm{O/H})$, stellar mass, star formation rate, and specific star formation rate, respectively. Values are adopted from the literature unless otherwise noted, with the corresponding references given in Column 9. \\\
    References:
(1) \citet{Berg2022};
(2) \citet{berg16};
(3) MPA-JHU\footnote{Data catalogs are available at \url{http://www.mpa-garching.mpg.de/SDSS/}. The Max Planck Institute for Astrophysics/Johns Hopkins University (MPA/JHU) SDSS database was produced by a collaboration of researchers (currently or formerly) from the MPA and JHU. The team includes Stephane Charlot (IAP), Guinevere Kauffmann and Simon White (MPA), Tim Heckman (JHU), Christy Tremonti (University of Wisconsin--Madison; formerly JHU), and Jarle Brinchmann (Leiden University; formerly MPA).};
(4) \citet{berg19a};
(5) \citet{senchyna22_civ};
(6) Determined from the reddening-corrected H$\beta$ flux measured from the SDSS spectrum, the luminosity distance, and the SFR calibration of \citet{kennicutt12};
(7) \citet{guseva04};
(8) \citet{corbin02}.}
\label{tbl:sample}
\end{deluxetable*}

\subsection{HST/COS FUV Spectroscopy}
The FUV spectra analyzed in this work were obtained with the Cosmic Origins Spectrograph (COS) on the \textit{Hubble Space Telescope} (HST), combining new observations with archival spectroscopy from several HST programs. We require observations with the medium-resolution G130M and G160M gratings ($R\gtrsim10{,}000$), which provide sufficient spectral resolution to separate the narrow nebular \ion{C}{4} components from the broader stellar-wind features and to resolve the detailed structure of the resonantly scattered \ion{C}{4} emission profiles. The full set of HST program IDs, PIs, datasets, gratings, central-wavelength settings, and exposure times is listed in Appendix~\ref{sec:observations}.

A principal source of the new spectroscopy is the Cycle 29 program HST-GO-16643 (PIs: Gazagnes and Berg), designed to obtain velocity-resolved Ly$\alpha$ and \ion{C}{4} profiles of nearby extreme emission-line galaxies selected for their unusually hard ionizing radiation fields. The observations were designed to achieve a continuum signal-to-noise ratio of S/N$>5$ near both Ly$\alpha$ and \ion{C}{4}. At the COS lifetime positions used for these observations, the nominal resolving power of the medium-resolution FUV gratings is $R\sim13{,}000$, corresponding to $\Delta v\sim23\,{\rm km\,s^{-1}}$. Because the targets are compact but not perfect point sources, their effective spectral resolution can be somewhat lower than the nominal instrumental resolution. Previous COS observations of similar compact star-forming galaxies indicate effective resolving powers as low as $R\sim8{,}000$ ($\Delta v\sim38\,{\rm km\,s^{-1}}$), which remains sufficient to resolve the \ion{C}{4} profile structure analyzed in this work.

\subsection{COS Reduction and Spectral Coaddition}
Five of the 18 galaxies are included in the COS Legacy Archive Spectroscopic SurveY \citep[CLASSY;][]{Berg2022,James2022}; for these targets, we adopt the high-resolution CLASSY G130M+G160M coadded spectra. For the remaining 13 galaxies, the COS observations were retrieved from MAST, processed with the standard CALCOS pipeline (version 3.4.4), and coadded following the CLASSY procedures \citep[described in][]{Berg2022,James2022} to ensure consistency across the full sample.

Briefly, individual exposures obtained with a given grating at different FP-POS positions were combined to reduce fixed-pattern noise. Exposures from different observing programs and central-wavelength settings were then aligned and coadded using exposure-time weighting. The resulting grating spectra were then placed on a common wavelength grid, with pixels affected by poor data-quality flags excluded or appropriately downweighted, and overlapping spectral regions were combined using the same exposure-time weighting scheme. The final spectra were binned by six native COS pixels per resolution element, corresponding to a velocity sampling of $\sim 30\,{\rm km\,s^{-1}}$ per binned element. This binning improves the S/N while preserving the resolved \ion{C}{4} emission and absorption structure. Finally, each spectrum was shifted to the galaxy systemic rest frame using the redshifts listed in Table~\ref{tbl:sample}, derived from the available optical spectra for the sample.

\subsection{Stellar Continuum Modeling and Normalization}\label{cont_fit}
Accurate treatment of the stellar continuum is particularly important for the \ion{C}{4}\,$\lambda\lambda1548,1550$ doublet because the observed profile can contain contributions from nebular emission, interstellar absorption, and broad P-Cygni features produced by winds from massive stars. We therefore model the FUV stellar continuum following the methodology of \citet{chisholm19} and \citet{Parker2026}. The stellar spectrum is represented as a luminosity-weighted linear combination of 30 single-age \texttt{Starburst99} \citep{leitherer99} simple stellar population models spanning three stellar metallicities, $Z_\star/Z_\odot=0.05$, 0.2, and 0.4, and ten population ages of 1, 2, 3, 4, 5, 8, 10, 15, 20, and 40 Myr. We restrict the models to these relatively low stellar metallicities to remain consistent with the low gas-phase abundances measured for the galaxies in our sample. The models assume a Kroupa initial mass function with an upper-mass cutoff of $100\,M_\odot$ and the high-mass-loss stellar evolutionary tracks of \citet{Meynet1994}. Each stellar population model also includes nebular continuum emission calculated with \texttt{Cloudy} photoionization models \citep{ferland13,ferland17}, assuming $n_e=100\,{\rm cm^{-3}}$ and $\log U=-2.5$. Because the \texttt{Starburst99} models have a fixed spectral resolution of 0.4\,\AA, the COS spectra are convolved to the same resolution for the stellar-population fitting.

The stellar continuum models were attenuated assuming a uniform foreground dust screen and the FUV attenuation curve of \citet{reddy16}. Strong nebular emission and interstellar absorption features, foreground contamination, and pixels affected by poor data quality were masked during the fit. The fitting windows were selected to retain broad stellar-wind and photospheric features that provide constraints on the age and metallicity of the massive-star population \citep[e.g.,][]{demello00,vidal-garcia17}, including the \ion{N}{5}\,$\lambda\lambda1238,1240$, \ion{Si}{4}\,$\lambda\lambda1393,1402$, and \ion{C}{4}\,$\lambda\lambda1548,1550$ P-Cygni wind features, while excluding the narrow nebular and interstellar components of these transitions. Broad \ion{He}{2}\,$\lambda1640$ emission was masked where necessary because the adopted \texttt{Starburst99} models do not adequately reproduce this feature in very young, metal-poor stellar populations. The fit was performed with \texttt{mpfit} \citep{markwardt09}, simultaneously solving for the light contributions of the individual stellar populations and the stellar reddening.

\begin{figure*}
\begin{center}
	\includegraphics[width=0.95\textwidth]{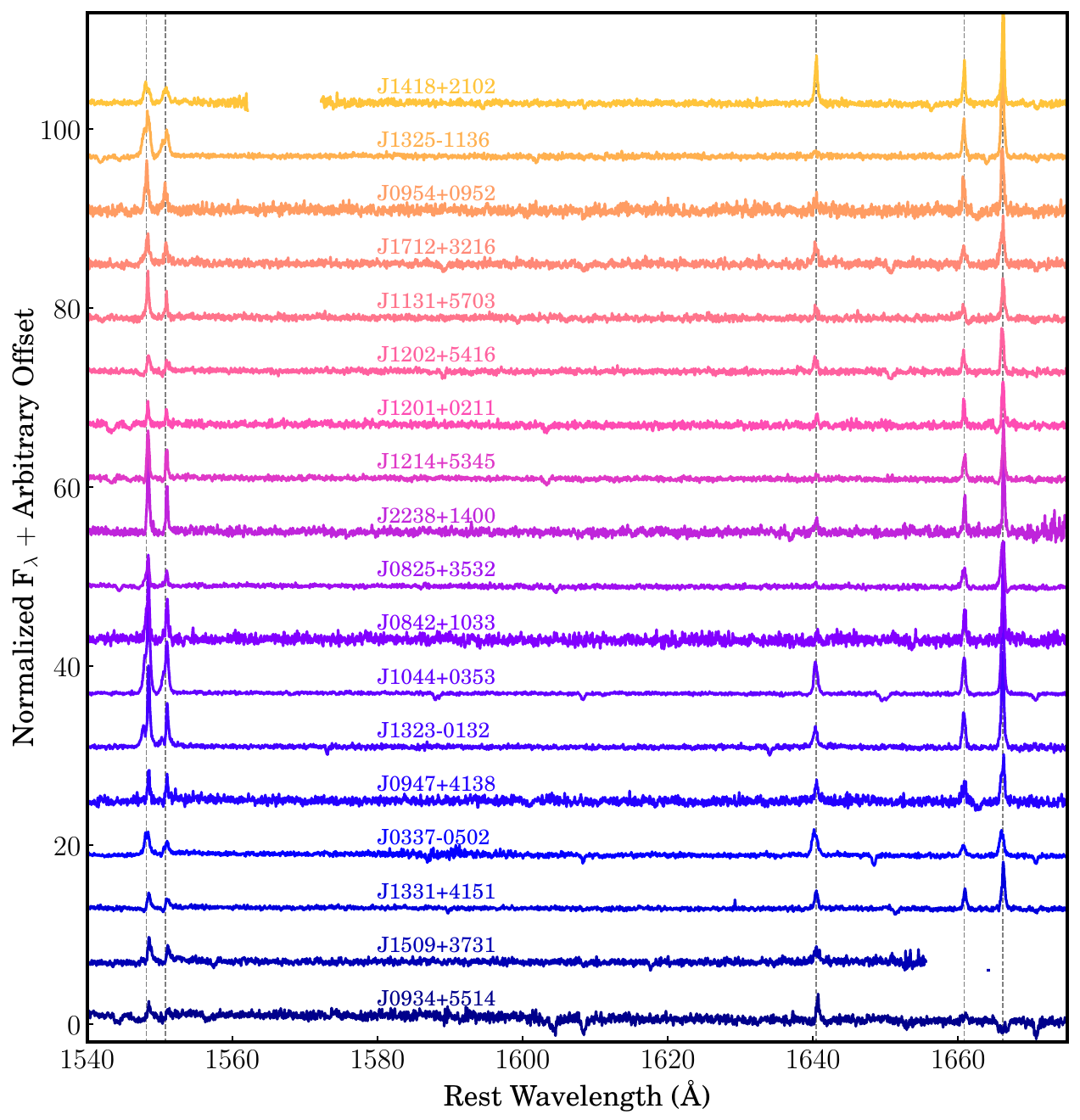}
\end{center}
\caption{\textbf{Continuum-normalized rest-frame UV spectra of the sample of 18 strong \ion{C}{4}-emitting galaxies used in this work.} The medium-resolution HST/COS spectra were normalized by the best-fitting \texttt{Starburst99}-based stellar population models, including nebular continuum emission, which also account for the broad stellar \ion{C}{4} P-Cygni contribution (see \S\ref{cont_fit}). The resulting spectra show a diversity of nebular \ion{C}{4} $\lambda\lambda1548,1550$ emission and resonant absorption profiles, together with strong high-ionization emission lines including \ion{He}{2} $\lambda1640$ and \ion{O}{3}] $\lambda\lambda1661,1666$, consistent with highly ionized gas exposed to hard radiation fields.}
\label{fig:civ_spectra}
\end{figure*}

For the \ion{C}{4} RT analysis, the original COS spectra were divided by their best-fitting stellar-plus-nebular continuum models. This normalization removes the broad wavelength-dependent continuum and corrects for the stellar \ion{C}{4} P-Cygni contribution, while preserving the narrow nebular \ion{C}{4} emission and the resonant absorption and re-emission produced by C$^{3+}$-bearing gas. The resulting continuum-normalized \ion{C}{4} spectra (shown in Figure~\ref{fig:civ_spectra}) are then compared with the continuum-normalized RT models in our subsequent analysis.

\section{\ion{C}{4} Radiative Transfer Model}
\label{sec:rt_model}

To interpret the resonant \ion{C}{4} profiles shown in Figure~\ref{fig:civ_spectra}, we present the first systematic application of the clumpy RT framework \texttt{PEACOCK} to observed \ion{C}{4}\,$\lambda\lambda1548,1550$ profiles in star-forming galaxies. The detailed numerical implementation is presented in \citet{Li2026a}; here, we summarize the physical ingredients most relevant for interpreting the observed line-profile morphology. These include the model geometry and gas kinematics, the continuum and intrinsic line-emission photon sources, and the treatment of aperture-dependent scattered emission.

\subsection{Model Geometry and Gas Kinematics}
\label{subsec:rt_geometry_kinematics}

We model \ion{C}{4} resonant scattering assuming spherical symmetry, with the UV-emitting source represented as a point source at the center of a clumpy circumgalactic medium (CGM). The gas is represented by a population of identical, spherical C$^{3+}$-bearing clumps distributed within a spherical halo bounded by an inner radius, $R_{\rm in}$, and an outer radius, $R_{\rm out}$. The clump number density declines radially as $n_{\rm cl}(r)\propto r^{-2}$. In our fiducial model, all clumps have the same radius and \ion{C}{4} column density, and we adopt $R_{\rm out}=10\,R_{\rm in}$.

The primary gas properties governing the shape of the emergent \ion{C}{4} profile are the line-of-sight \ion{C}{4} column density, $N_{\rm CIV,\,LOS}$; the internal Doppler broadening of individual clumps, $b_{\rm D,\,cl}$; the macroscopic velocity dispersion of the clump population, $\sigma_{\rm cl}$; and the bulk radial velocity field of the clumps. The latter is parameterized by a radially varying clump outflow velocity profile characterized by a maximum velocity, $v_{\rm out,\,max}$ (see Equations 1--3 of \citealt{Li2026a}). As shown below in Section \ref{sec:rt_experiments}, the bulk outflow of the clumps plays a particularly important role in producing red--blue asymmetries in the emergent \ion{C}{4} profile. As the outflow velocity increases, line photons preferentially escape on the red side of the transition, while continuum photons can produce blueshifted absorption troughs. This behavior is analogous to the formation of asymmetric double-peaked and P-Cygni-like profiles in other resonant transitions, such as Ly$\alpha$ \citep{Li2026a}.

\subsection{Photon Sources: Continuum and Line Emission}
\label{subsec:rt_sources}

In principle, the observed \ion{C}{4} profile can contain contributions from both continuum photons and intrinsically produced nebular \ion{C}{4} line photons, with both components subsequently modified by resonant scattering through C$^{3+}$ gas clumps. We therefore include two types of input photon sources: a continuum source and an intrinsic line-emission source. Importantly, the RT model does not include any in situ production of \ion{C}{4} photons within the outflowing gas, such as collisionally excited nebular emission. Thus, any \ion{C}{4} emission associated with the wind in the model arises entirely from resonant scattering and redistribution of the input continuum and intrinsic line photons. In addition, since the observed spectra are divided by the best-fitting stellar-plus-nebular continuum model before the RT analysis (Section~\ref{cont_fit}), the continuum source is taken to be flat in the continuum-normalized spectra. 

For the continuum-only simulations, photons are injected with a flat spectrum over a velocity interval $\Delta V$ surrounding the \ion{C}{4} doublet. The corresponding wavelength interval is
\begin{equation}
\Delta \lambda = \lambda_0 \frac{\Delta V}{c}
\end{equation}
where $\lambda_0$ is the mean wavelength of the \ion{C}{4} doublet.

For the line-only simulations, photons are injected at the two \ion{C}{4} transitions at 1548.20~\AA\ and 1550.78~\AA, with each component modeled as a Gaussian centered at the corresponding line center and having an intrinsic velocity dispersion of $\sigma_{\rm int}=25\,{\rm km\,s^{-1}}$. Their intrinsic emissivity ratio is fixed to
\begin{equation}
\frac{j_{1548}}{j_{1550}} = 2
\end{equation}
where $j_{1548}$ and $j_{1550}$ denote the line-integrated intrinsic emissivities. This ratio approximates the expectation for collisionally excited \ion{C}{4} emission in the low-density, optically thin limit.

Because the RT calculation is linear in photon number, the continuum-only and line-only simulations can be combined with arbitrary relative weights to construct composite spectra. We define $R_{\rm line}$ as the ratio of the total number of intrinsically emitted \ion{C}{4} line photons to the number of continuum photons emitted over the wavelength interval $\Delta\lambda$

\begin{equation}
R_{\rm line}
\equiv
\frac{N_{\rm line}}
     {N_{\rm cont}(\Delta\lambda)}
=
\frac{N_{\rm line}}
     {n_{\lambda,{\rm cont}}\,\Delta\lambda}
\end{equation}
where $N_{\rm line}$ is summed over both components of the \ion{C}{4} doublet and $n_{\lambda,{\rm cont}}$ is the continuum photon number density per unit wavelength. The intrinsic equivalent width (EW) of the combined doublet
is therefore
\begin{equation}
{\rm EW}_{\rm int}
\equiv
\frac{N_{\rm line}}{n_{\lambda,{\rm cont}}}
=
R_{\rm line}\,\Delta\lambda
=
R_{\rm line}\,\frac{\lambda_0\Delta V}{c}
\end{equation}

In the illustrative RT models below, we adopt $\Delta V=1500\,{\rm km\,s^{-1}}$, wide enough to encompass both transitions of the doublet. 

\subsection{Aperture Treatment}
\label{subsec:rt_aperture}

A key feature of resonant line transfer is that photons can scatter to large projected distances before escaping. Consequently, the observed line profile depends not only on the intrinsic emission and the physical properties of the scattering medium, but also on the fraction of spatially extended scattered light captured by the observational aperture. As we show below, this aperture effect is particularly important for \ion{C}{4}, because scattered emission emerging at large impact parameters can partially fill in absorption troughs and alter the relative strengths of the blue and red emission components.

To model this effect, we record the escape impact parameter $b$ of each photon, defined as the projected distance between the escaping photon trajectory and the center of the halo. Spectra can then be constructed either by selecting photons within specified ranges of the normalized impact parameter, $b/R_{\rm out}$, or cumulatively by including all photons with $b\leq b_{\rm max}$. In the latter case, the dimensionless quantity $b_{\rm max}/R_{\rm out}$ serves as an effective aperture parameter, with larger values corresponding to the recovery of photons scattered to progressively larger projected distances from the central source. This parameter should not be interpreted as a direct one-to-one measurement of the physical COS aperture. Instead, it provides a controlled way to quantify how the emergent \ion{C}{4} morphology changes as different fractions of spatially extended scattered emission are included. 

\section{Physical Origin of \ion{C}{4} Profile Morphologies}
\label{sec:rt_experiments}

Before fitting the observed spectra, we use a set of idealized RT experiments to build physical intuition for the diversity of \ion{C}{4} profiles produced by resonant scattering. These experiments are intended to isolate how the emergent \ion{C}{4} morphology depends on the \ion{C}{4} column density, gas kinematics, intrinsic \ion{C}{4} equivalent width, and aperture-dependent recovery of spatially extended scattered emission. This section therefore provides a physical basis for interpreting the profile classes and best-fit models presented in later sections.

\subsection{\ion{C}{4} Line Profiles in Radial Annuli}
\label{subsec:radial_binned_profiles}

We first examine how resonant scattering redistributes \ion{C}{4} photons in both frequency and projected radius. Figure~\ref{fig:non_outflowing} presents the emergent spectra in multiple radial annuli for two limiting source models: an intrinsic line-only source and a continuum-only source. Each line-only simulation uses $4\times10^4$ photons, while each continuum-only simulation uses $10^5$ photons. These photon numbers were chosen based on convergence tests: fewer photons produce noticeably larger numerical noise in the emergent spectra, whereas increasing the photon numbers further produces negligible improvement while substantially increasing the computational cost. The larger photon number in the continuum-only simulations compensates for the broader wavelength range over which continuum photons are injected. Escaping photons are grouped by their projected escape impact parameter, $b$, with the colored curves showing spectra from individual radial annuli and the black curve showing the spectrum integrated over all escaping photons.

\subsubsection{Non-outflowing Case}

\begin{figure*}
\begin{center}
	\includegraphics[width=\textwidth]{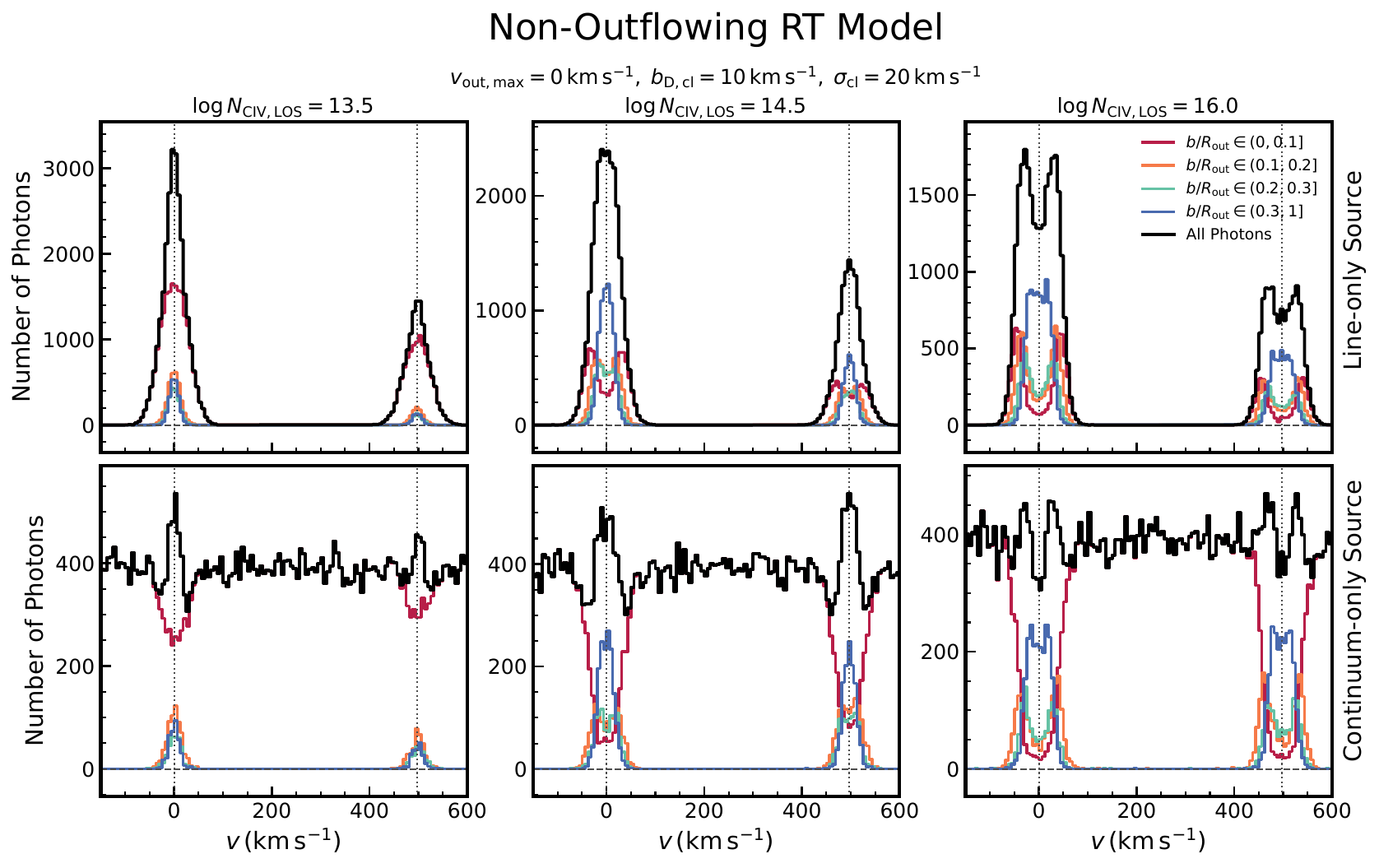} 
\end{center}
\caption{
\textbf{Radially resolved \ion{C}{4} profiles for the non-outflowing clumpy RT model.} The upper row shows the line-only source model, while the lower row shows the continuum-only source model. From left to right, the LOS \ion{C}{4} column density increases from
$\log(N_{\rm CIV,\,LOS}/{\rm cm}^{-2})=13.5$ to 14.5 and 16.0.
Colored curves group escaping photons by their projected escape impact parameter, $b/R_{\rm out}$:
$(0,0.1]$, $(0.1,0.2]$, $(0.2,0.3]$, and $(0.3,1.0]$,
while the black curve includes all escaping photons.
Here, $b$ is the projected distance from the halo center at which an escaping photon is observed, normalized by the outer radius $R_{\rm out}$. Increasing \ion{C}{4} column density enhances both frequency and spatial diffusion: line-photon profiles evolve toward stronger central absorption and symmetric double-peaked structure, while continuum photons are increasingly removed from central regions and re-emitted at larger impact parameters.
}
\label{fig:non_outflowing}
\end{figure*}

To isolate the basic effects of resonant scattering, we first consider models with no bulk outflow. Figure~\ref{fig:non_outflowing} shows the resulting profiles for three total line-of-sight (LOS) \ion{C}{4} column densities, increasing from left to right: $\log(N_{\rm CIV,\,LOS}/{\rm cm}^{-2})=13.5$, 14.5, and 16.0. The top row shows the line-only source models, while the bottom row shows the continuum-only source models. In all cases, each clump has a Doppler parameter of $b_{\rm D,\,cl}=10\,{\rm km\,s^{-1}}$, accounting for both internal thermal and non-thermal motions, and the clump population has a macroscopic velocity dispersion of $\sigma_{\rm cl}=20\,{\rm km\,s^{-1}}$.

In the absence of bulk outflow, increasing the LOS \ion{C}{4} column density primarily increases the line-center optical depth and enhances the frequency diffusion of resonantly scattered photons. The top row of Figure~\ref{fig:non_outflowing} shows that the line-only spectra consequently evolve from single-peaked profiles at $\log(N_{\rm CIV,\,LOS}/{\rm cm}^{-2})=13.5$ (left panel), to profiles with stronger line-center absorption at $\log(N_{\rm CIV,\,LOS}/{\rm cm}^{-2})=14.5$ (middle panel), and finally to clearly double-peaked profiles at $\log(N_{\rm CIV,\,LOS}/{\rm cm}^{-2})=16.0$ (right panel). 

The same increase in optical depth also redistributes escaping photons spatially, producing systematic changes among the radial annuli. At $\log(N_{\rm CIV,\,LOS}/{\rm cm}^{-2})=13.5$ in the top-left panel of Figure~\ref{fig:non_outflowing}, more than half of the line photons escape through the innermost annulus, $b/R_{\rm out}\in(0,0.1]$ (red curve), and the integrated profile (black curve) remains relatively compact and single-peaked. At larger impact parameters, the $(0.1,0.2]$ (orange), $(0.2,0.3]$ (green), and $(0.3,1.0]$ (blue) annuli show broadly similar profile shapes but substantially lower photon counts than the innermost annulus. At $\log(N_{\rm CIV,\,LOS}/{\rm cm}^{-2})=14.5$ in the top-middle panel, resonant scattering redistributes a larger fraction of photons to greater impact parameters. The inner annuli ($b/R_{\rm out} \lesssim 0.3$) begin to develop stronger line-center absorption and double-peaked structure, while the outermost annulus contributes an increasingly large fraction of the escaping photons. The integrated profile therefore remains approximately single-peaked. At $\log(N_{\rm CIV,\,LOS}/{\rm cm}^{-2})=16.0$ in the top-right panel, the line-center absorption becomes much deeper and the peak separation increases relative to the lower-column-density models in all radial annuli. The outermost annulus remains an important contributor to the spatially extended emission and itself develops pronounced frequency structure, while the inner annuli contribute sufficiently strongly that the integrated spectrum acquires a clear central trough and becomes double-peaked.

\begin{figure*}
\begin{center}
	\includegraphics[width=\textwidth]{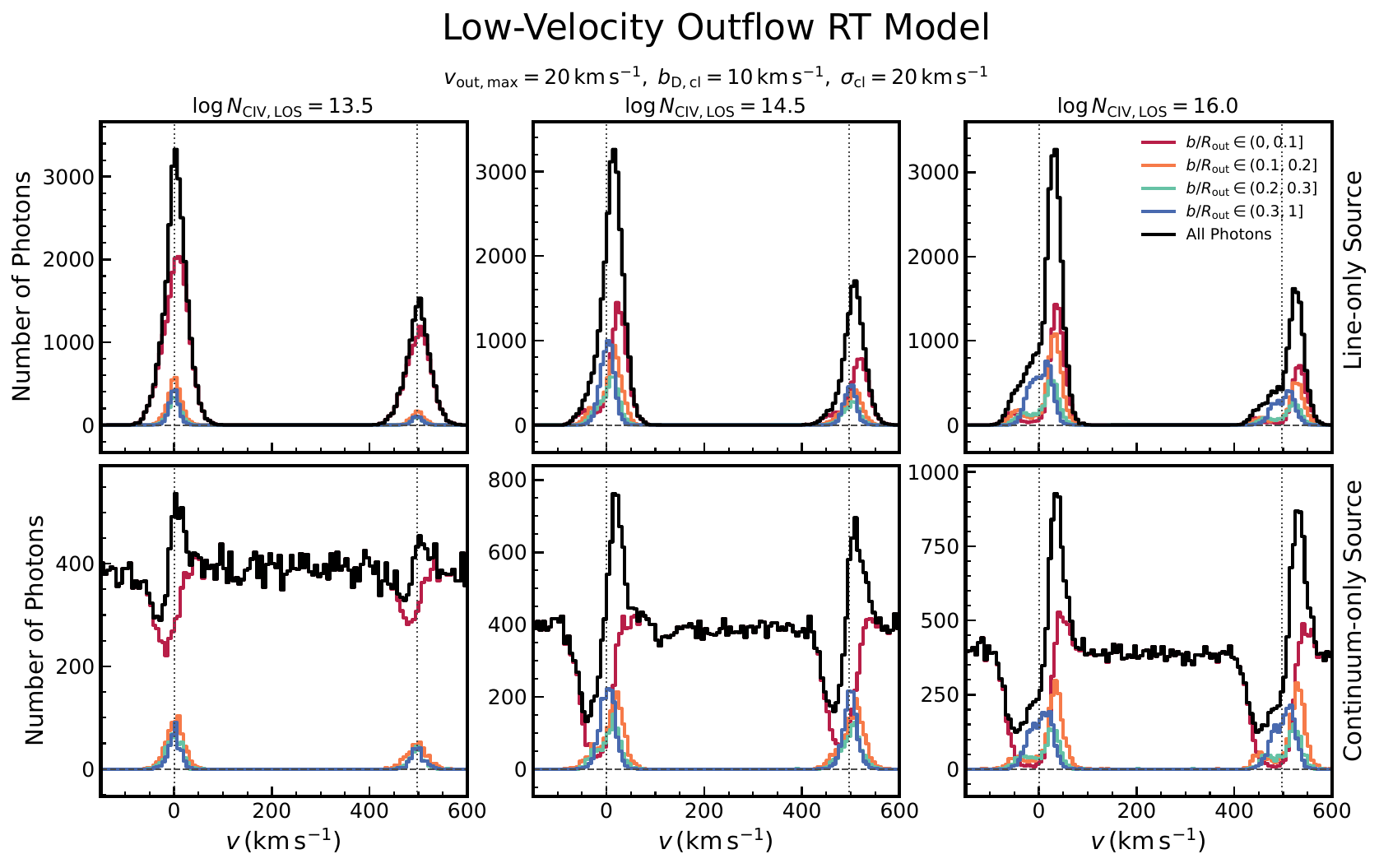}  
\end{center}
\caption{
\textbf{Radially resolved \ion{C}{4} profiles for the low-velocity outflow RT model.} The upper row shows simulations with line-only photon sources, while the lower row shows continuum-only photon sources. From left to right, the LOS \ion{C}{4} column density increases from $\log(N_{\rm CIV,\,LOS}/{\rm cm}^{-2})=13.5$ to 14.5 and 16.0, with $v_{\rm out,\,max}=20\,{\rm km\,s^{-1}}$ in all panels. Colored curves show photons escaping through different projected radial annuli, $b/R_{\rm out}\in(0,0.1]$, $(0.1,0.2]$, $(0.2,0.3]$, and $(0.3,1.0]$, while the black curve shows the profile integrated over all escaping photons. Here, $b$ is the projected escape impact parameter normalized by the outer radius, $R_{\rm out}$. Even a modest bulk outflow breaks the red--blue symmetry of the non-outflowing models, producing increasingly red-dominated emission in the line-only source models and P-Cygni-like profiles with blueshifted absorption and redshifted scattered emission in the continuum-only source models.
}
\label{fig:outflowing_v20}
\end{figure*}

The continuum-only source models in the second row of Figure \ref{fig:non_outflowing} show a complementary radial pattern. Continuum photons near each \ion{C}{4} resonance are scattered out of the central region and re-emitted preferentially at larger impact parameters. At $\log(N_{\rm CIV,\,LOS}/{\rm cm}^{-2})=13.5$ in the bottom-left panel, the innermost annulus (red curve) is therefore dominated by resonant absorption, whereas the outer annuli show scattered emission. Their combination produces an integrated profile with a central emission peak bracketed by absorption troughs. At $\log(N_{\rm CIV,\,LOS}/{\rm cm}^{-2})=14.5$ in the bottom-middle panel, the central absorption deepens and extends into the intermediate annuli as the optical depth increases. Finally, at $\log(N_{\rm CIV,\,LOS}/{\rm cm}^{-2})=16.0$ in the bottom-right panel, line-center absorption is present across all radial annuli. The integrated central emission feature then separates into a double-peaked structure, with the line-center trough falling below the continuum level.

\subsubsection{Outflowing Cases}

We next introduce a radially varying bulk outflow velocity for the clumps, beginning with a modest maximum velocity of $v_{\rm out,\,max}=20\,{\rm km\,s^{-1}}$. As shown in Figure~\ref{fig:outflowing_v20}, even this relatively slow outflow breaks the red--blue symmetry of the static models and produces systematic asymmetries in the line profiles of both line-only and continuum-only source models.

\begin{figure*}
\begin{center}
	\includegraphics[width=\textwidth]{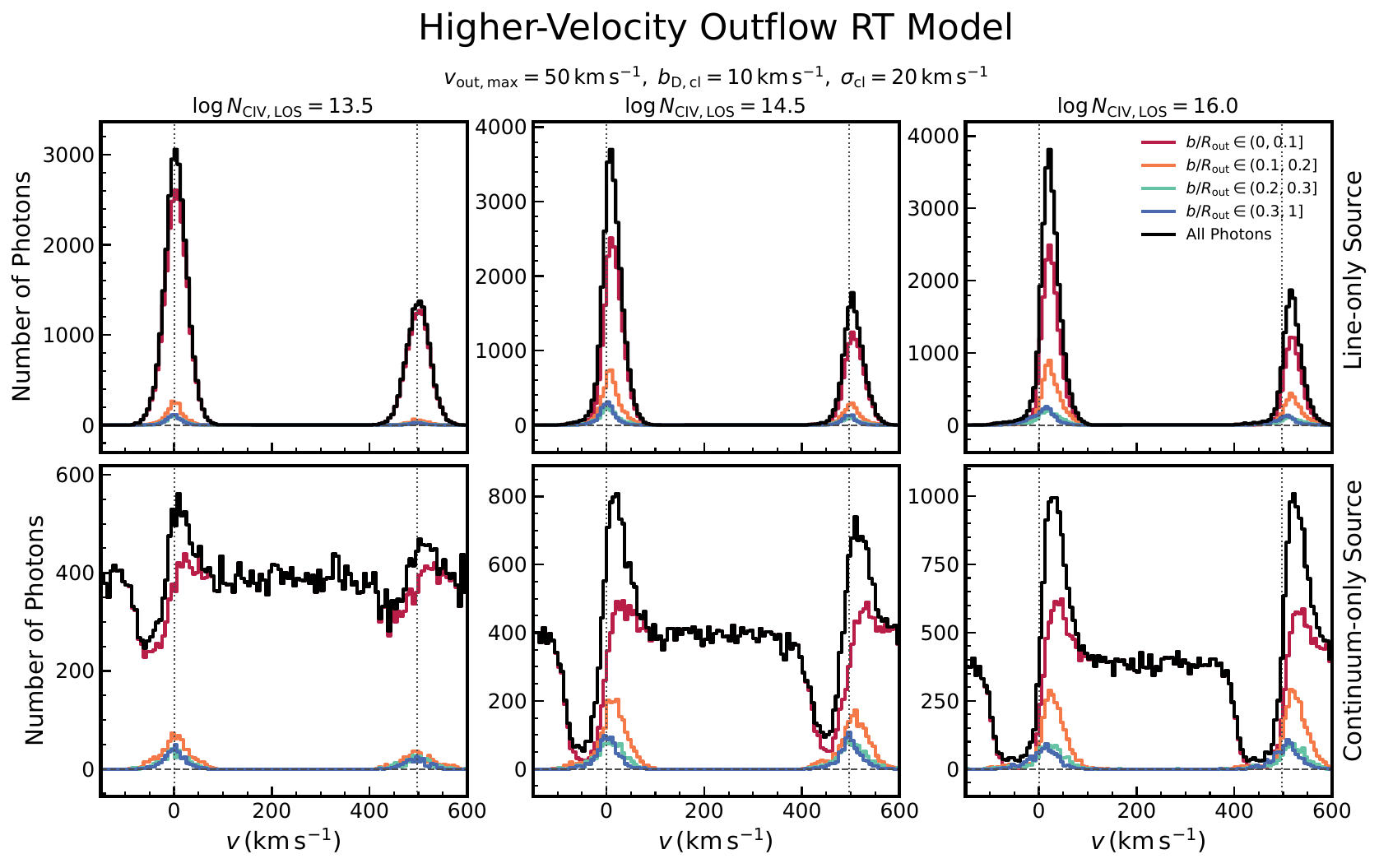}    
\end{center}
\caption{
\textbf{Radially resolved \ion{C}{4} profiles for the higher-velocity outflow RT model.} The upper row shows simulations with line-only photon sources, while the lower row shows continuum-only photon sources. From left to right, the LOS \ion{C}{4} column density increases from $\log(N_{\rm CIV,\,LOS}/{\rm cm}^{-2})=13.5$ to 14.5 and 16.0, with $v_{\rm out,\,max}=50\,{\rm km\,s^{-1}}$ in all panels. Colored curves show photons escaping through different projected radial annuli, $b/R_{\rm out}\in(0,0.1]$, $(0.1,0.2]$, $(0.2,0.3]$, and $(0.3,1.0]$, while the black curve shows the profile integrated over all escaping photons. Here, $b$ is the projected escape impact parameter normalized by the outer radius, $R_{\rm out}$. Compared with the low-velocity outflow model, the stronger bulk motion further enhances the red--blue asymmetry, suppressing the blue emission component and strengthening red-dominated profiles in the line-only source models, while producing more strongly blueshifted absorption and redshifted scattered emission in the continuum-only source models.
}
\label{fig:outflowing_v50}
\end{figure*}

The top row of Figure~\ref{fig:outflowing_v20} shows the line-only source models. At $\log(N_{\rm CIV,\,LOS}/{\rm cm}^{-2})=13.5$ in the left panel, the inner annuli already start to show a mild red--blue asymmetry. At $\log(N_{\rm CIV,\,LOS}/{\rm cm}^{-2})=14.5$ in the middle panel, this asymmetry becomes more pronounced with a weaker blue ``bump'' and a stronger red peak, and significant line-center absorption begins to emerge in the inner annuli. By $\log(N_{\rm CIV,\,LOS}/{\rm cm}^{-2})=16.0$ in the right panel, the central absorption is substantially deeper and the blue and red peaks are more clearly separated. Increasing column density also redistributes a larger fraction of photons to greater impact parameters. In particular, the outermost annulus, $b/R_{\rm out}\in(0.3,1.0]$ (blue curve), develops a prominent scattered component that contributes a blue bump to the integrated spectrum. Overall, the outflowing line-only models are dominated by redshifted emission, whose velocity shifts increasingly redward as $N_{\rm CIV,\,LOS}$ increases.

The bottom row of Figure~\ref{fig:outflowing_v20} shows that the continuum-only source model develops a P-Cygni-like morphology, characterized by a blueshifted absorption trough with a redshifted emission peak. At $\log(N_{\rm CIV,\,LOS}/{\rm cm}^{-2})=13.5$ in the left panel, the innermost annulus, $b/R_{\rm out}\in(0,0.1]$, shows strong blueshifted absorption together with a mild redshifted emission component, producing a P-Cygni-like profile. The outer annuli are dominated by scattered emission, which adds to the redshifted emission in the integrated spectrum. At $\log(N_{\rm CIV,\,LOS}/{\rm cm}^{-2})=14.5$ in the middle panel, the blueshifted absorption strengthens and the scattered emission from the intermediate and outer annuli becomes more pronounced. By $\log(N_{\rm CIV,\,LOS}/{\rm cm}^{-2})=16.0$ in the right panel, the extended scattered components become increasingly redshifted and develop more complex, double-peaked structure. These photons partially fill the absorption trough in the integrated spectrum. The flux minimum in the innermost annulus remains near $\sim-20\,{\rm km\,s^{-1}}$, close to $-v_{\rm out,\,max}$ where the radial optical depth is largest. In contrast, the minimum of the integrated absorption trough is shifted to larger blueshifted velocities, reaching $\sim-50\,{\rm km\,s^{-1}}$, as spatially extended re-emission preferentially fills in the absorption closer to line center.

To demonstrate the effect of a stronger outflow, we further increase the maximum outflow velocity to $v_{\rm out,\,max}=50\,\rm km\,s^{-1}$ in Figure~\ref{fig:outflowing_v50}. At fixed \ion{C}{4} column density, the larger outflow velocity further increases the red--blue asymmetry, shifting scattered emission farther to the red and absorption farther to the blue in both source models. In the top row of Figure~\ref{fig:outflowing_v50}, the blue bump present in the lower-velocity outflow model is strongly suppressed in the line-only source model, while the emergent profile becomes increasingly dominated by the red peak for the same set of other model parameters. The stronger outflow also changes the spatial distribution of escaping photons: a larger fraction escapes at smaller impact parameters, especially in the $b/R_{\rm out}\in(0,0.1]$ and $(0.1,0.2]$ annuli, while the outermost annulus, $b/R_{\rm out}\in(0.3,1.0]$, contributes less to the total escaped flux. Notably, although the red peak becomes more dominant, its velocity does not shift substantially farther to the red. This suggests that increasing $v_{\rm out,max}$ primarily changes the red-to-blue flux ratio and the spatial redistribution of photons, rather than simply translating the entire profile to higher velocity.

In the bottom row of Figure~\ref{fig:outflowing_v50}, the innermost annulus in the continuum-only source model retains its strong blueshifted absorption but develops a more prominent redshifted scattered-emission component than in the lower-velocity outflow model. The absorption trough shifts further blueward, with its flux minimum reaching $\sim-50\,{\rm km\,s^{-1}}$. The outer annuli remain dominated by scattered emission but become increasingly red-peak dominated, similar to the line-only source model. Because fewer photons are re-emitted on the blue side, the scattered-emission infilling of the blueshifted absorption trough becomes much weaker. As a result, the integrated profile more closely follows the innermost annulus, producing a deeper and more nearly saturated blueshifted absorption trough.

\subsection{Composite \texorpdfstring{C~{\sc iv}}{CIV} Model Line Profiles}\label{sec:composite_models}

Having examined the line-only and continuum-only source models separately, we now combine the two components to construct composite spectra that more closely resemble observed \ion{C}{4} line profiles. Each composite model consists of an intrinsic nebular line component and a flat continuum component. Their relative contribution to the composite input spectrum is characterized by the intrinsic equivalent width, ${\rm EW}_{\rm int}$. For a given ${\rm EW}_{\rm int}$, we first convert the emergent line-only and continuum-only spectra into spectra per injected photon. We then keep the continuum normalization fixed and rescale the line component according to
\begin{equation}
\frac{N_{\rm line}}{N_{\rm cont}} =
\frac{{\rm EW}_{\rm int}}{\Delta\lambda}
\end{equation}
where $\Delta\lambda = \lambda_0 {\Delta V}/{c}$ and $\Delta V$ is the velocity interval over which the continuum is defined. The final composite spectrum is obtained by adding the fixed continuum spectrum and the rescaled line spectrum. Such a reweighting scheme avoids the stochastic noise introduced by photon subsampling and isolates the effect of changing the intrinsic line-to-continuum ratio on the emergent \ion{C}{4} profile. For comparison across different models and apertures, we normalize the resulting composite spectra by the continuum level, $F_{\rm cont}$, and present the line profiles in units of $F/F_{\rm cont}$.

\begin{figure*}
\begin{center}
	\includegraphics[width=0.78\textwidth]{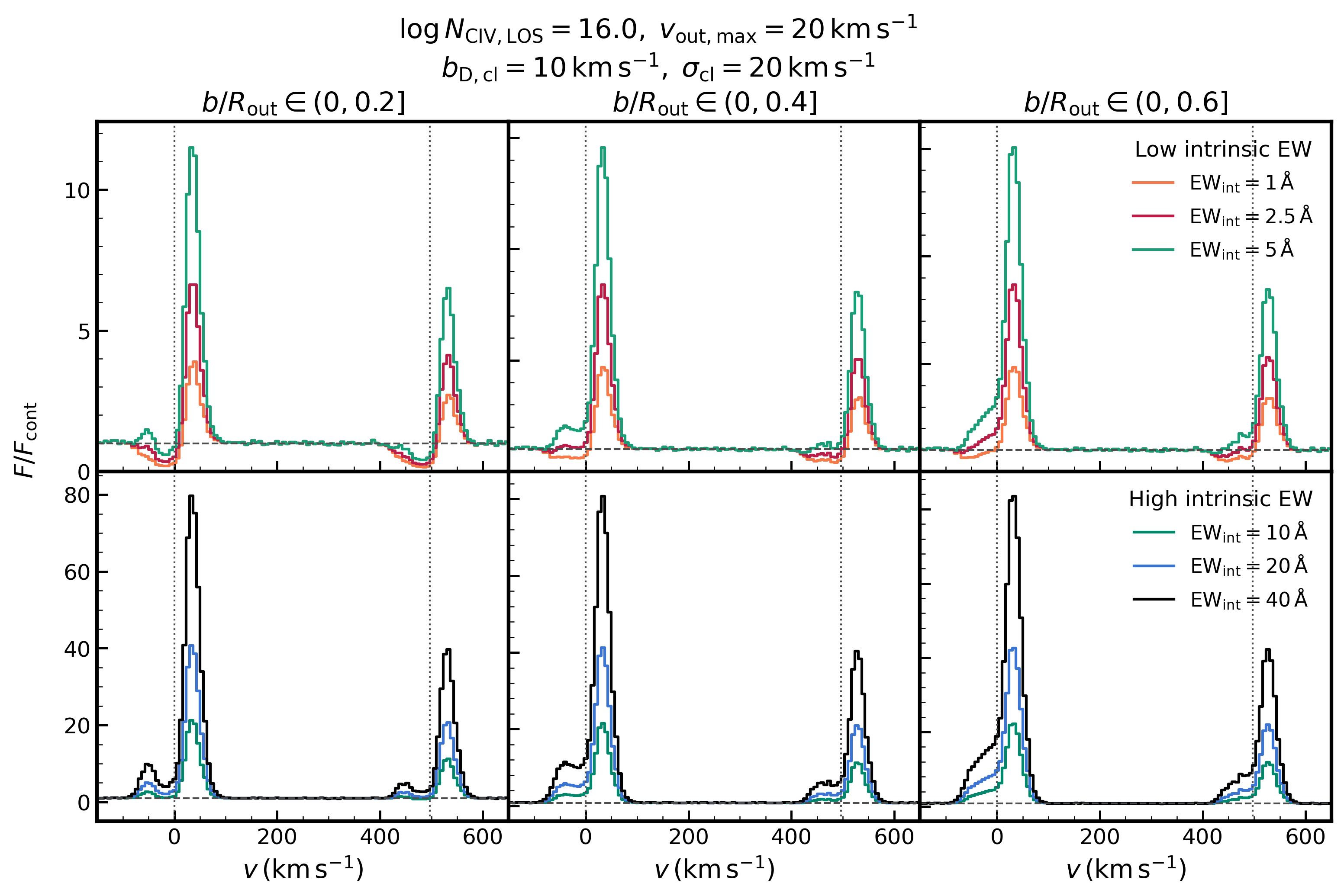}  \\
    \includegraphics[width=0.78\textwidth]{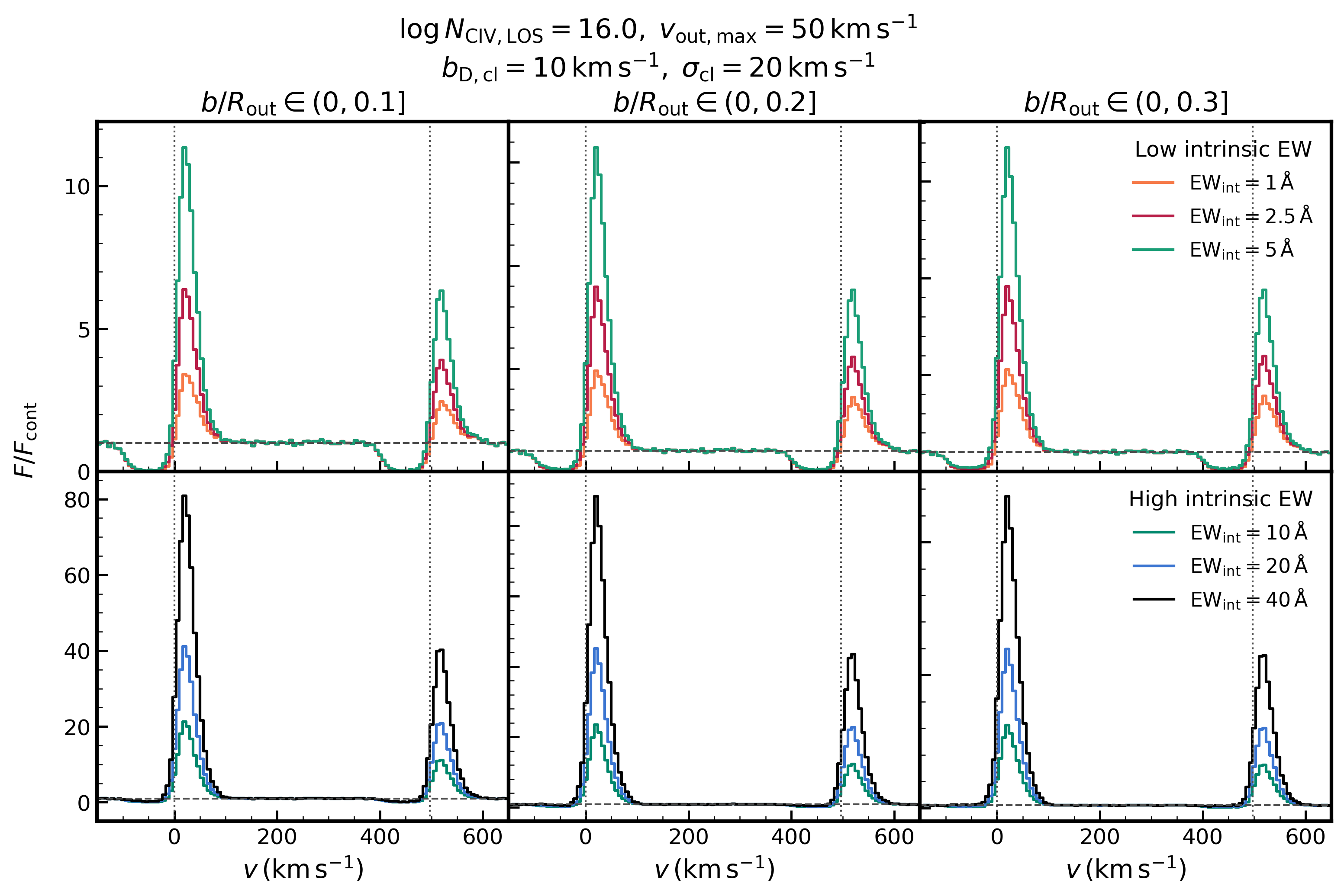}  
\end{center}
\caption{\textbf{Composite \ion{C}{4} line profiles for different intrinsic line-to-continuum ratios.} The models are constructed by combining the emergent line-only and continuum-only spectra after rescaling the line component to the specified intrinsic equivalent width, ${\rm EW}_{\rm int}$. From top to bottom, the two sets of panels show $(\log (N_{\rm CIV,\,LOS}/{\rm cm^{-2}}), v_{\rm out,\,max}/{\rm km\,s^{-1}})=(16.0,20)$ and $(16.0,50)$. Within each set, the columns correspond to different normalized escape impact parameter ranges, $b/R_{\rm out}$, which mimic different effective observational apertures. Curves of different colors show different ${\rm EW}_{\rm int}$ values, and all spectra are normalized by the continuum level. These models produce the four representative profile types discussed in the text: P-Cygni-like profiles, double-peaked emission profiles, double-peaked profiles with two absorption troughs, and ``blue-bump + red-peak'' profiles.}
\label{fig:composite_models}
\end{figure*}

Figure~\ref{fig:composite_models} shows several composite \ion{C}{4} line profiles for two high column density models with $\log (N_{\rm CIV,\,LOS}/{\rm cm^{-2}})=16.0$. The upper set corresponds to a low outflow velocity model with $v_{\rm out,\,max}=20\,{\rm km\,s^{-1}}$, while the lower set shows a faster outflow model with $v_{\rm out,\,max}=50\,{\rm km\,s^{-1}}$. In each case, we vary the intrinsic equivalent width, ${\rm EW}_{\rm int}$, and extract spectra within different maximum normalized impact parameters, $b/R_{\rm out}$, which mimic different effective observational apertures. The ${\rm EW}_{\rm int}$ values explored here span the range observed in extreme \ion{C}{4} emitters across cosmic time. Typical rest-frame \ion{C}{4} EWs are $\sim1$--$10\,{\rm \AA}$ at $z\sim0$ \citep[e.g.,][]{Senchyna2019,Berg2021}, $5$ -- $30\,{\rm \AA}$ at $z\sim3$ -- 5 \citep[e.g.,][]{Saxena2022,mascia23_civvandels}, $10$ -- $40\,{\rm \AA}$ at $z\sim6$ -- 9 \citep[e.g.,][]{Stark2015,Topping2025}, and $20$ -- $50\,{\rm \AA}$ at $z>9$ \citep[e.g.,][]{Castellano2024,Tang2026}. The lower-${\rm EW}_{\rm int}$ models therefore overlap with local and intermediate-redshift extreme emitters, while the largest ${\rm EW}_{\rm int}$ values probe the high-EW regime reported at the highest redshifts.

To guide the discussion, we describe four representative composite-profile morphologies, summarized in Table~\ref{tab:types}: Type A, P-Cygni-like profiles; Type B, double-peaked emission profiles; Type C, double-peaked profiles with two absorption troughs; and Type D, ``blue-bump + red-peak'' profiles. These types are not intended to provide an exhaustive classification of all possible \ion{C}{4} line profiles, nor should they be interpreted as one-to-one diagnostics of any single physical parameter. Instead, they are used here as descriptive labels for the combined effects of \ion{C}{4} column density, outflow velocity, intrinsic line strength, and aperture-dependent recovery of spatially extended scattered emission.

The upper set of panels in Figure~\ref{fig:composite_models} shows the $v_{\rm out,\,max}=20\,{\rm km\,s^{-1}}$ model, for which the morphology changes significantly with both ${\rm EW}_{\rm int}$ and aperture. For the smallest aperture, $b/R_{\rm out}\in(0,0.2]$, low-${\rm EW}_{\rm int}$ spectra remain continuum-dominated and show a P-Cygni-like morphology (Type A). As ${\rm EW}_{\rm int}$ increases, the blueshifted absorption trough becomes progressively shallower as the red emission peaks strengthen relative to the continuum, and the profile transitions toward a double-peaked shape (Type B). At intermediate aperture, $b/R_{\rm out}\in(0,0.4]$, the red curve with ${\rm EW}_{\rm int} = 2.5\,{\rm \AA}$ provides a clear example of a Type C profile, with two emission peaks and two apparent absorption troughs, one of which lies near the line center, while the other is strongly blueshifted. Importantly, this highly blueshifted trough does not directly trace the bulk outflow velocity; instead, it arises because scattered emission partially fills the original absorption trough. For the largest aperture, $b/R_{\rm out}\in(0,0.6]$, extended blue-side scattered emission becomes more prominent, especially at large ${\rm EW}_{\rm int}$, producing a ``blue-bump + red-peak'' morphology (Type D).

\begin{table*}
\centering
\caption{Representative morphological classes of composite \ion{C}{4} line profiles}
\label{tab:types}
\noindent\hspace*{-25.74pt}\makebox[\textwidth][c]{\resizebox{\textwidth}{!}{%
\begin{tabular}{lll}
\hline
\hline
Type & Morphology & Physical Scenario \\
\hline
A & P-Cygni & Continuum-dominated profile; strong resonant absorption with limited emission infilling \\
B & Double peaks & Line-emission-dominated profile with residual absorption between the two peaks \\
C & Double peaks + double troughs & Moderate $\rm EW_{\rm int}$; partial infilling creates an additional blueshifted trough \\
D & Blue bump + red peak & Large-aperture recovery of extended blueshifted scattered emission \\
\hline
\end{tabular}}}\par
\end{table*}

\begin{figure*}
\begin{center}
    \includegraphics[width=0.9\textwidth]{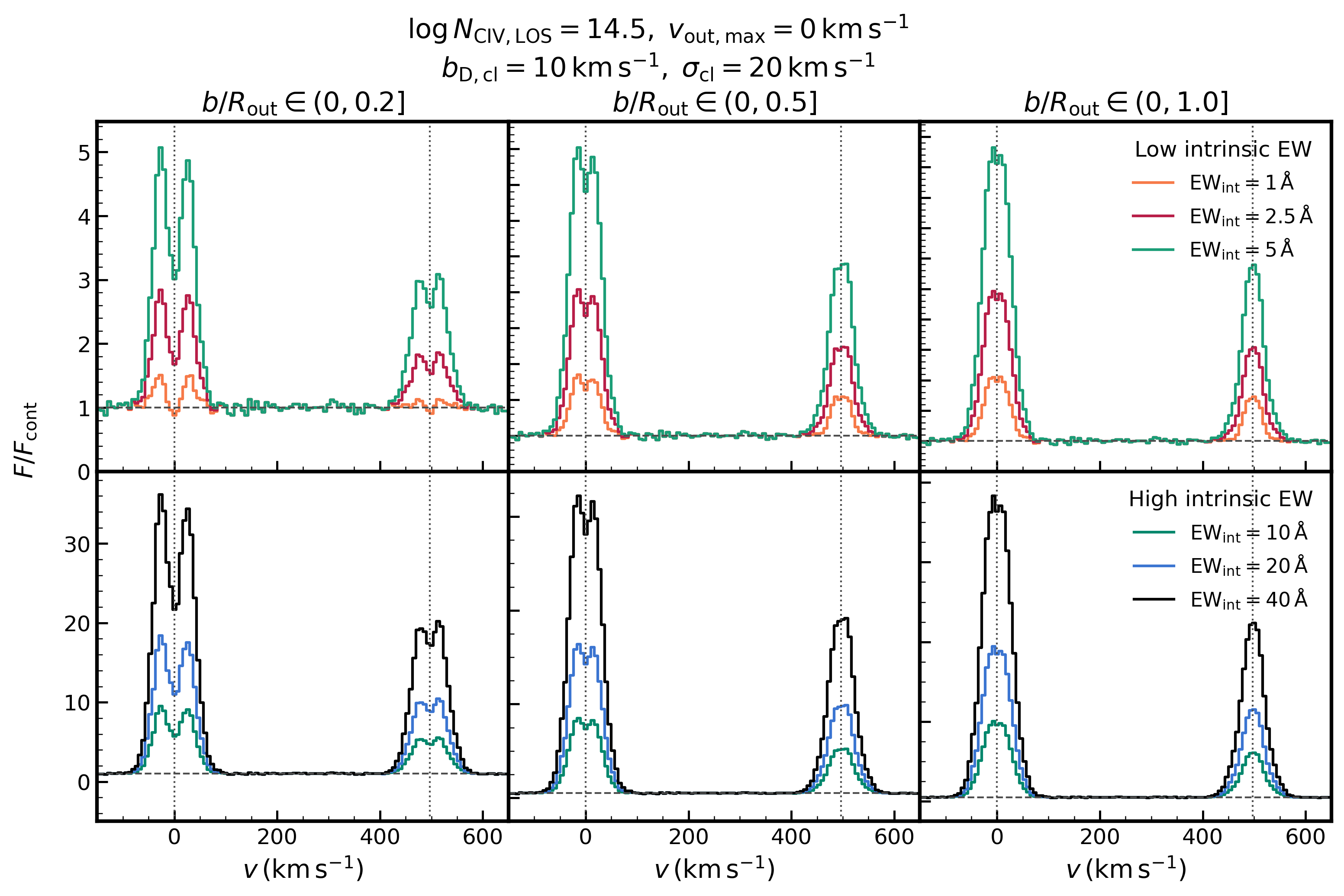}  \\
\end{center}
\caption{\textbf{Composite \ion{C}{4} model line profiles for a moderate column density, non-outflowing model.} The model has $\log (N_{\rm CIV,\,LOS}/{\rm cm^{-2}})=14.5$ and $v_{\rm out,\,max}=0\,{\rm km\,s^{-1}}$. Different columns show spectra extracted within different normalized escape impact parameters, $b/R_{\rm out}$, mimicking different effective observational apertures. The profiles are approximately symmetric about the two \ion{C}{4} doublet components and show double-peaked structure, especially at small apertures. As the aperture increases, resonantly scattered photons escaping at larger impact parameters contribute more strongly near line center, producing emission infilling that weakens the central absorption features. }
\label{fig:composite_models2}
\end{figure*}

The lower set of panels in Figure~\ref{fig:composite_models} shows the same column density but a faster outflow, with $v_{\rm out,\,max}=50\,{\rm km\,s^{-1}}$. The larger outflow velocity shifts the scattered emission preferentially toward the red side and weakens the blue-side emission infilling. It also shifts photons out of resonance more efficiently, reducing the number of scatterings and hence the amount of spatial diffusion before escape. As a result, the escaping photons are more concentrated at small impact parameters, and the composite profiles show little additional evolution once the aperture reaches $b/R_{\rm out}\sim0.3$. We therefore show a smaller aperture range, $b/R_{\rm out}\in(0,0.1]$ -- $(0,0.3]$, for this model. Across this range of ${\rm EW}_{\rm int}$ and $b/R_{\rm out}$, the spectra remain closer to P-Cygni than in the $v_{\rm out,\,max}=20\,{\rm km\,s^{-1}}$ model. Even at high ${\rm EW}_{\rm int}$, the blueshifted absorption trough remains prominent, with a flux minimum located close to $-v_{\rm out,\,max}\simeq -50\,{\rm km\,s^{-1}}$. In addition, the profiles do not develop the same strong blue-bump morphology seen in the low-velocity case. This comparison demonstrates that ${\rm EW}_{\rm int}$ alone does not uniquely determine the observed \ion{C}{4} profile shape; the velocity structure of the scattering medium and the aperture-dependent recovery of extended emission are equally important.

Figure~\ref{fig:composite_models2} shows a moderate column density, non-outflowing model with $\log (N_{\rm CIV,\,LOS}/{\rm cm^{-2}})=14.5$ and $v_{\rm out,\,max}=0\,{\rm km\,s^{-1}}$. In contrast to the high column density outflowing models discussed above, the profiles are nearly symmetric about the two \ion{C}{4} doublet components. For the smallest aperture, $b/R_{\rm out}\in(0,0.2]$, the composite spectra show clear double-peaked structure around each doublet component, with residual flux near line center increasing as ${\rm EW}_{\rm int}$ becomes larger. As the aperture is increased to $b/R_{\rm out}\in(0,0.5]$ and $(0,1.0]$, photons scattered to larger impact parameters contribute more strongly near line center, partially filling the central absorption features (cf. Figure~\ref{fig:non_outflowing}). The resulting profiles become smoother and can resemble intrinsic double-Gaussian emission profiles that have not been significantly modified by RT effects. This resemblance, however, is misleading: the apparent double-Gaussian morphology is actually produced by the superposition of line-center absorption at small impact parameters and scattered line-center emission from larger impact parameters. 

\begin{figure*}
\begin{center}
	\includegraphics[width=\textwidth]{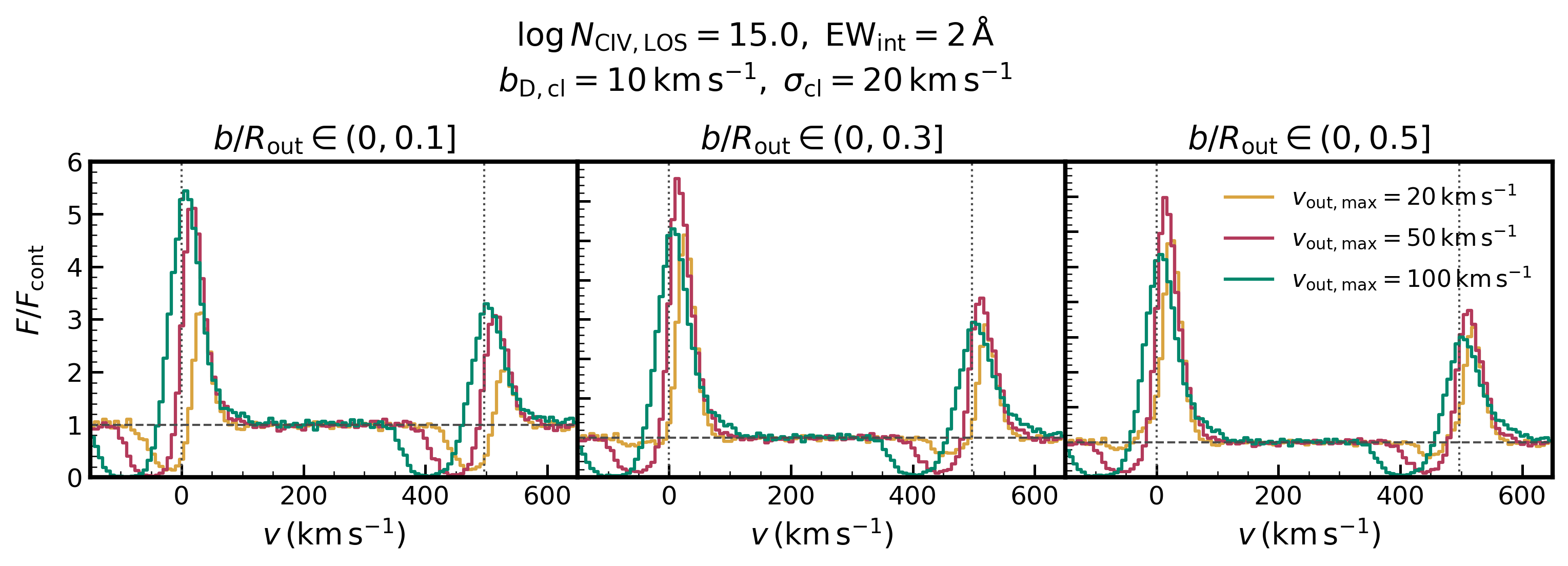}\\
    \includegraphics[width=\textwidth]{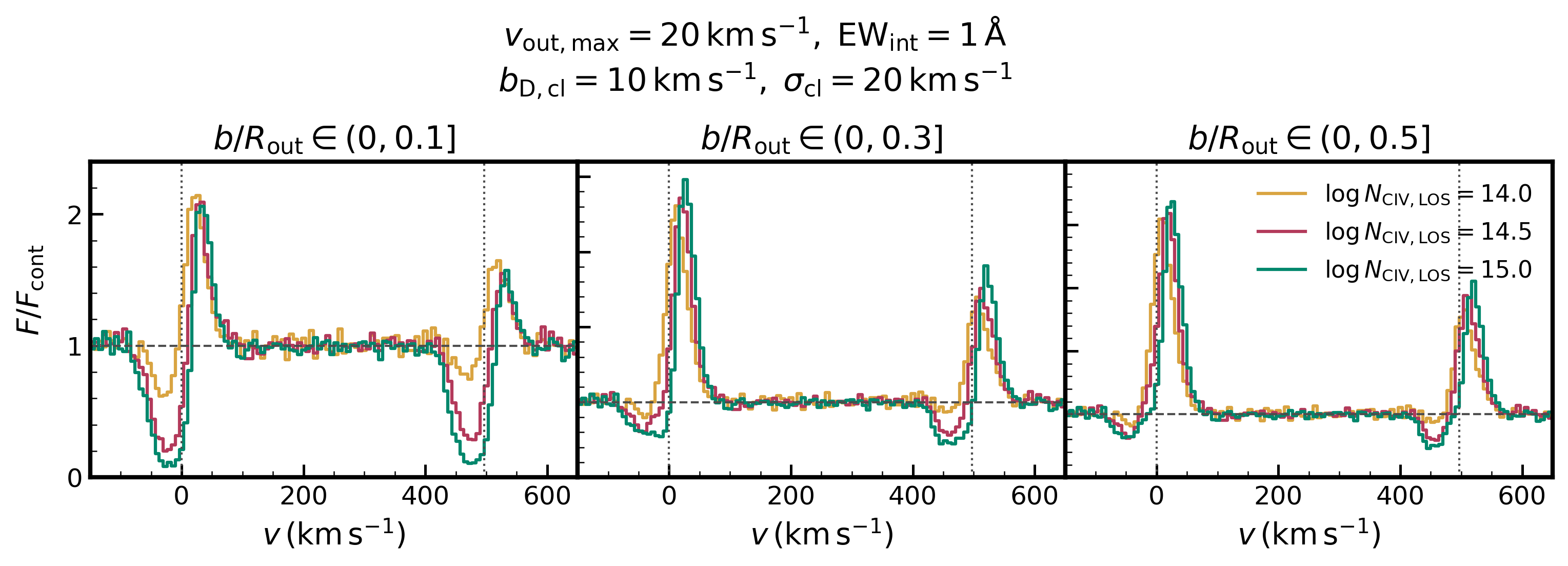}     
\end{center}
\caption{\textbf{Dependence of composite \ion{C}{4} line profiles on outflow velocity and LOS column density.} The spectra are constructed using the same line-rescaling method as in Figure~\ref{fig:composite_models2} and are normalized by the continuum level, $F_{\rm cont}$. In the upper set of panels, we fix $\log (N_{\rm CIV,\,LOS}/{\rm cm^{-2}})=15.0$ and ${\rm EW}_{\rm int}=2\,{\rm \AA}$, and vary the maximum outflow velocity, $v_{\rm out,\,max}$. In the lower set of panels, we fix $v_{\rm out,\,max}=20\,{\rm km\,s^{-1}}$ and ${\rm EW}_{\rm int}=1\,{\rm \AA}$, and vary $\log (N_{\rm CIV,\,LOS}/{\rm cm^{-2}})$. In both sets, columns show increasing $b/R_{\rm out}$. Increasing $v_{\rm out,\,max}$ shifts the blueshifted absorption trough to more negative velocities, while increasing $N_{\rm CIV,\,LOS}$ strengthens resonant absorption and enhances frequency diffusion. Larger apertures increase the contribution of spatially extended scattered emission, modifying the apparent trough depth and shape, while leaving the red emission peak velocity comparatively less affected.}
\label{fig:varying_vmax_N}
\end{figure*}

Having shown that composite \ion{C}{4} line profile morphology is shaped by a complex interplay among the RT parameters, intrinsic EW, and effective aperture, we now perform an additional set of experiments in which we vary $v_{\rm out,\,max}$ and $N_{\rm CIV,\,LOS}$ and see how the line profile changes accordingly. In the upper row of Figure~\ref{fig:varying_vmax_N}, we fix $\log (N_{\rm CIV,\,LOS}/{\rm cm^{-2}})=15.0$ and ${\rm EW}_{\rm int}=2\,{\rm \AA}$, and vary $v_{\rm out,\,max}$ from $20$ to $100\,{\rm km\,s^{-1}}$. In the lower row, we instead fix $v_{\rm out,\,max}=20\,{\rm km\,s^{-1}}$ and ${\rm EW}_{\rm int}=1\,{\rm \AA}$, and vary $\log (N_{\rm CIV,\,LOS}/{\rm cm^{-2}})$ from 14.0 to 15.0. In both cases, the three columns show increasing $b/R_{\rm out}$.

In the velocity sequence shown in the upper set of panels, increasing $v_{\rm out,\,max}$ primarily shifts the blueshifted absorption trough to more negative velocities. This behavior is clearest for the smallest aperture, $b/R_{\rm out}\in(0,0.1]$, where the profile is dominated by continuum photons passing through the inner halo and is only weakly affected by spatially extended scattered emission. In this regime, the velocity of the absorption minimum approximately traces $-v_{\rm out,\,max}$, the velocity at which the optical depth of the outflowing clumps is largest. The red emission peaks also change with outflow velocity: as $v_{\rm out,\,max}$ increases, the location of the red peak moves blueward and becomes closer to the line center.

As $b/R_{\rm out}$ increases, scattered emission from larger impact parameters contributes more strongly to the total line profile. This effect is most pronounced in the lowest-velocity outflow model, $v_{\rm out,\,max}=20\,{\rm km\,s^{-1}}$, where blue-side scattered emission partially fills and blurs the blueshifted absorption trough. When the aperture is increased to $b_{\rm max}/R_{\rm out}\in(0,0.5]$, this infilling shifts the apparent minimum of the blue trough in the $v_{\rm out,\,max}=20\,{\rm km\,s^{-1}}$ model to approximately $-50\,{\rm km\,s^{-1}}$. Importantly, although increasing $b/R_{\rm out}$ at fixed $v_{\rm out,\,max}$ modifies the shape and apparent depth of the absorption trough, the velocity of the red emission peak is comparatively less affected by aperture.

The column-density sequence in the lower set of panels shows a complementary effect. At fixed $v_{\rm out,\,max}=20\,{\rm km\,s^{-1}}$, increasing $N_{\rm CIV,\,LOS}$ raises the resonant optical depth and strengthens the blueshifted absorption. For the smallest $b/R_{\rm out}$, the red emission peak shifts toward larger positive velocities, the blueshifted absorption trough becomes deeper, whereas the velocity of the trough minimum remains largely unchanged. As $b/R_{\rm out}$ increases, scattered emission increasingly fills the troughs, which shifts the apparent flux minimum to more negative velocities. The effect is strongest for the two higher column density models. As in the velocity-sequence models shown in the top row, the velocity of the red emission peak remains largely unchanged as $b/R_{\rm out}$ is increased.

Overall, these experiments show that different spectral features respond to different physical and observational parameters, including the \ion{C}{4} column density, gas kinematics, intrinsic \ion{C}{4} EW, and effective observing aperture. For instance, the velocity of the blueshifted absorption minimum is primarily controlled by the outflow velocity, but its apparent location can be biased significantly by aperture-dependent emission infilling. The red emission peak, by contrast, is more sensitive to the \ion{C}{4} column density and the velocity structure of the scattering medium, and is comparatively insensitive to the adopted aperture. These results demonstrate that composite \ion{C}{4} profiles cannot be interpreted using a single spectral feature or integrated line strength alone. Robust interpretation requires RT modeling that simultaneously accounts for intrinsic \ion{C}{4} production, resonant scattering through C$^{3+}$ gas clumps, gas kinematics, and the fraction of spatially extended scattered emission recovered in the observed spectrum.

\subsection{Conditions for Minimal Emission Infilling of Blueshifted Absorption}
\label{sec:minimal_infilling}

An important consideration (particularly for observers) in interpreting blueshifted \ion{C}{4} absorption is the extent to which the absorption trough is filled in by resonantly scattered emission. Our RT experiments identify several physical regimes in which this effect is minimized.

Ideally, spatially resolved observations can isolate the central region of the outflow. Spectra extracted at small apertures are least affected by emission infilling because they exclude a large fraction of the resonantly scattered photons that emerge at larger impact parameters (see Figures~\ref{fig:non_outflowing}--\ref{fig:outflowing_v50}). However, when only spatially integrated observations are available and the observing aperture encompasses a substantial fraction of the extended emission, emission infilling can still remain relatively insignificant under two conditions. First, for sufficiently strong outflows (e.g., $v_{\rm out,\,max}\gtrsim50\,{\rm km\,s^{-1}}$ in our models), the scattered line emission is strongly shifted toward the red side of the profile. Consequently, relatively little scattered emission falls within the blueshifted absorption trough, reducing the degree of blue-side infilling (see Figure~\ref{fig:varying_vmax_N}, upper row). Second, at relatively low \ion{C}{4} column densities (e.g., $\log(N_{\rm CIV,\,LOS}/{\rm cm^{-2}})\lesssim14.0$), resonant scattering is less effective at producing extended emission at large projected radii. In particular, the extended blue emission component becomes weak, such that spectra integrated over larger apertures experience less infilling of the blueshifted absorption trough (see Figure~\ref{fig:varying_vmax_N}, lower row).

In summary, for spatially integrated observations, emission infilling is expected to be less significant in systems with sufficiently large outflow velocities (as indicated by strongly blueshifted absorption troughs) and/or relatively low \ion{C}{4} column densities (as suggested by unsaturated absorption). Conversely, slowly outflowing and optically thicker systems observed with large apertures are expected to be the most susceptible to emission infilling. In such cases, it becomes difficult to infer physical parameters directly from the observed spectral signatures, highlighting the need for RT modeling to properly account for emission infilling.

\subsection{Intrinsic and Observed Equivalent Widths}
\label{sec:ewint_vs_ewobs}

\begin{figure*}
\begin{center}
	\includegraphics[width=\textwidth]{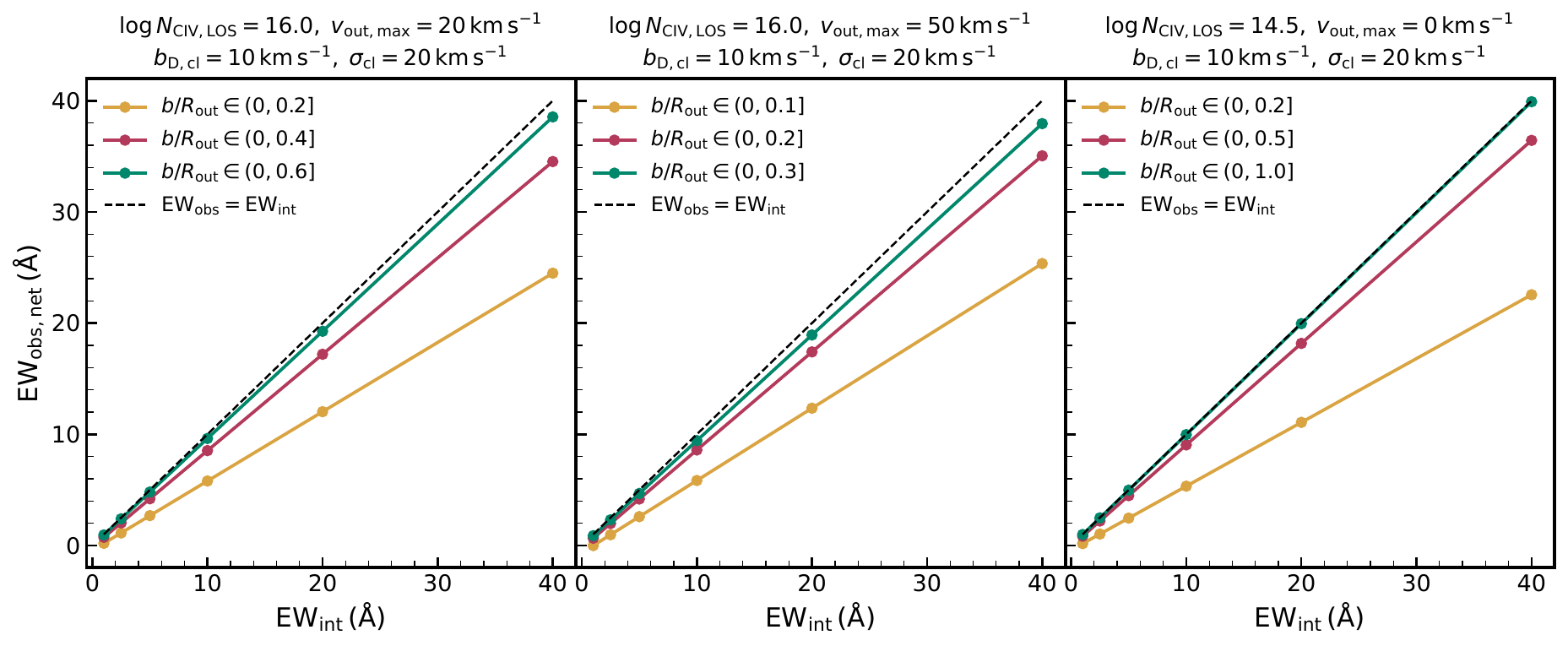}    
\end{center}
\caption{\textbf{Comparison between intrinsic and observed net equivalent widths in composite \ion{C}{4} RT models.}
The three panels show representative dust-free models with different LOS \ion{C}{4} column densities and outflow velocities, as labeled above each panel. Colored curves show ${\rm EW}_{\rm obs,\,net}$ measured from the aperture-dependent composite spectra as a function of the imposed intrinsic equivalent width, ${\rm EW}_{\rm int}$, for different maximum normalized impact parameters, $b/R_{\rm out}$. The black dashed line marks the one-to-one relation, ${\rm EW}_{\rm obs,\,net}={\rm EW}_{\rm int}$. In general, ${\rm EW}_{\rm obs,\,net}$ is smaller than ${\rm EW}_{\rm int}$, especially for small apertures that miss a larger fraction of spatially extended scattered emission. As $b/R_{\rm out}$ increases, more scattered line photons are recovered and the measured net EW approaches the intrinsic value.}
\label{fig:ewint_vs_ewobs}
\end{figure*}

The intrinsic equivalent width used in the composite models, ${\rm EW}_{\rm int}$, sets the relative strength of the two input photon components, nebular line emission and continuum emission, before any radiative transfer. By contrast, the observed net equivalent width, ${\rm EW}_{\rm obs,\,net}$, is measured from the final aperture-dependent composite profile, including both emission peaks and absorption troughs
\begin{equation}
{\rm EW}_{\rm obs,\,net}
=
\int \left(\frac{F_\lambda}{F_{\rm cont}} - 1\right)\,d\lambda
\end{equation}
which means ${\rm EW}_{\rm obs,\,net}$ is not simply a source property; it also depends on how resonant scattering redistributes photons in frequency and projected radius. Comparing ${\rm EW}_{\rm int}$ with ${\rm EW}_{\rm obs,net}$ therefore provides a useful way to connect the emergent spectrum to the intrinsic balance between line and continuum emission.

Figure~\ref{fig:ewint_vs_ewobs} compares ${\rm EW}_{\rm obs,\,net}$ and ${\rm EW}_{\rm int}$ for the three representative composite models discussed above. All three models shown here are dust-free; if dust were included in the \ion{C}{4} clumps, ${\rm EW}_{\rm obs,\,net}$ would likely be further reduced because a fraction of the resonantly scattered line photons would be absorbed by dust before escaping. In general, the recovered net EW is smaller than the imposed intrinsic EW, especially for small values of $b/R_{\rm out}$. This offset arises because small apertures recover only a fraction of the spatially extended scattered line emission, while continuum absorption troughs still contribute negative flux to the net EW. As a result, the same intrinsic line-to-continuum photon ratio can produce a substantially smaller observed net EW when the aperture misses a significant fraction of the scattered emission.

The offset between ${\rm EW}_{\rm obs,\,net}$ and ${\rm EW}_{\rm int}$ decreases as $b/R_{\rm out}$ increases. This trend is visible in the first two panels of Figure~\ref{fig:ewint_vs_ewobs}: larger apertures include more resonantly scattered photons escaping at large projected radii, causing the measured net EW to move closer to the intrinsic value. In the limit where the aperture captures nearly all escaping line photons, the models approach ${\rm EW}_{\rm obs,\,net}\simeq{\rm EW}_{\rm int}$. This behavior is most clearly seen in the non-outflowing, moderate-column-density model in the third panel, where the largest aperture, $b/R_{\rm out}\in(0,1.0]$, nearly follows the one-to-one relation. More generally, the observed ${\rm EW}_{\rm obs,\,net}$ should be interpreted as a lower limit on the intrinsic ${\rm EW}_{\rm int}$. Systematic RT modeling is therefore essential for inferring the underlying intrinsic EW, especially if ${\rm EW}_{\rm int}$ is to be further translated into physical source properties through additional photoionization modeling.

\section{Fitting Observed \texorpdfstring{C~{\sc iv}}{CIV} Profiles}\label{sec:fitting_CIV_data}

Having developed a physical understanding of \ion{C}{4} RT and the formation of various line-profile morphologies, we now apply the clumpy RT framework to the observed HST/COS \ion{C}{4}\,$\lambda\lambda1548,1550$ profiles. This section describes the overall fitting strategy, the forward modeling of the observed spectra, the neural-network-accelerated Bayesian inference procedure, and the resulting best-fit profiles and morphological classes.

\subsection{Fitting Strategy}
\label{subsec:fitting_strategy}

The goal of the fitting is to reproduce the full velocity-resolved \ion{C}{4} doublet morphology, rather than only the integrated flux or equivalent width. For each galaxy, we compare the continuum-normalized observed spectrum with a customized library of composite RT models, each including both intrinsic \ion{C}{4} line emission and continuum photons that undergo resonant scattering through C$^{3+}$ gas clumps. The fitted parameters therefore describe both intrinsic \ion{C}{4} production and the properties of the scattering medium, including its column density, kinematics, and effective aperture-dependent scattered emission.

Since direct Monte Carlo RT calculations are computationally expensive, we use a neural-network emulator trained on the customized RT model library for each object. The emulator is then coupled to a nested-sampling algorithm to explore the posterior distribution of the model parameters and identify the best-fitting solution.

\subsection{Forward Modeling of the Observed Spectra}
\label{subsec:forward_model}

For each target, we shift the continuum-normalized COS spectrum to the systemic rest frame and construct a velocity grid relative to the \ion{C}{4}\,$\lambda1548$ transition. The model spectra include both transitions of the \ion{C}{4} doublet at rest wavelengths 1548.20~\AA\ and 1550.78~\AA. Spectral pixels affected by foreground absorption, detector artifacts, or unrelated features are masked when necessary.

For each RT model, we generate a set of emergent spectra by selecting escaped photons within different maximum impact parameters, $b_{\rm max}/R_{\rm out}$. This post-processing step mimics different effective observational apertures and controls the fraction of spatially extended scattered emission included in the model spectrum. Because the galaxies in our sample span a range of redshifts and physical sizes, the fixed 2.5$^{\prime\prime}$ COS aperture subtends different physical areas and captures different fractions of the \ion{C}{4}-emitting region from target to target. Each aperture-selected model spectrum, sampled at $16\,\mathrm{km\,s^{-1}}$, is shifted by the fitted velocity offset and then rebinned onto the observed $32\,\mathrm{km\,s^{-1}}$ velocity grid by averaging the model flux density over each observed velocity bin. The resulting profile is compared with the data for likelihood evaluation. In the fitting, $b_{\rm max}/R_{\rm out}$ is treated as an effective aperture parameter and is varied together with the physical RT parameters.

\subsection{Neural-network Emulator and Bayesian Inference}
\label{subsec:dnn_nested}

\renewcommand{\arraystretch}{1.15}
\begin{table*}
\footnotesize
\centering
\caption{Summary of RT model and fitting parameters used for the \ion{C}{4} profile modeling}
\label{tab:model_parameters}
\noindent\hspace*{-25.18pt}\makebox[\textwidth][c]{\resizebox{\textwidth}{!}{%
\begin{tabular}{lll}
\hline
\hline
Symbol & Parameter & Role in the Model \\
\hline
$N_{\rm CIV,\,LOS}$ 
& LOS \ion{C}{4} column density 
& Controls the resonant optical depth and frequency diffusion of \ion{C}{4} photons \\

$v_0$ & Outflow velocity normalization & Characteristic velocity scale of the clump radial outflow  \\

$R$ & Acceleration parameter & Dimensionless ratio of outward acceleration to gravitational deceleration \\

$b_{\rm D,\,cl}$ 
& Clump Doppler parameter 
& Describes the internal thermal \& non-thermal velocity dispersion within clumps \\

$\sigma_{\rm cl}$ 
& Clump velocity dispersion 
& Describes the macroscopic random velocity dispersion among clumps \\

$\tau_{\rm d,\,cl}$ & Dust optical depth & Effective dust absorption optical depth of each clump \\

${\rm EW}_{\rm int}$ 
& Intrinsic \ion{C}{4} equivalent width 
& Sets the relative contribution of intrinsic \ion{C}{4} line emission and continuum photons \\

$b_{\rm max}/R_{\rm out}$ 
& Effective aperture parameter 
& Controls the fraction of scattered emission included in the model spectrum \\

$\Delta v$ 
& Velocity offset 
& Accounts for small offsets between the model and observed systemic velocity frames \\

$f_{\rm scale}$ 
& Flux scaling factor 
& Accounts for small normalization differences between the model and observed spectra \\
\hline
\end{tabular}}}\par
\end{table*}

For each galaxy, we construct a customized library of Monte Carlo RT models spanning the parameter space relevant for the observed \ion{C}{4} profile. The main RT and fitting parameters are summarized in Table~\ref{tab:model_parameters}. The primary varied parameters include the LOS \ion{C}{4} column density, $N_{\rm CIV,\,LOS}$; the outflow velocity normalization, $v_0$; the acceleration parameter, $R$;\footnote{$v_0$ and $R$ jointly determine the maximum outflow velocity, $v_{\rm out,\,max}$  \citep{Li2026a}.} the clump Doppler parameter, $b_{\rm D,\,cl}$; the macroscopic clump velocity dispersion, $\sigma_{\rm cl}$; the clump dust optical depth, $\tau_{\rm d,\,cl}$; the intrinsic \ion{C}{4} equivalent width, ${\rm EW}_{\rm int}$; and the effective aperture parameter, $b_{\rm max}/R_{\rm out}$. In addition, we include two nuisance parameters in the spectral fitting: a velocity offset, $\Delta v$, to account for small residual uncertainties in the systemic velocity frame, and a flux scaling factor, $f_{\rm scale}$, to account for small normalization differences between the model and observed spectra.

Our fitting pipeline follows \citet{Li2026a}. For each object, we first construct a customized mock spectral library consisting of 200 Monte Carlo RT models spanning the parameter space relevant for the observed \ion{C}{4} profile. We then train a deep neural network (DNN) emulator to learn the mapping between the RT model parameters and the corresponding continuum-normalized model spectra. For each trial point in the parameter space, the DNN-predicted spectrum is compared with the observed \ion{C}{4} profile over the adopted fitting window using a Gaussian likelihood

\begin{equation}
\ln \mathcal{L}
=-\frac{1}{2}\sum_{i}
\left[
\frac{\left(F_i^{\rm obs}-F_i^{\rm mod}(\boldsymbol{x})\right)^2}
{\sigma_{{\rm tot},\,i}^2}
+
\ln\left(2\pi\sigma_{{\rm tot},\,i}^2\right)
\right]
\end{equation}
where ${\sigma_{{\rm tot},\,i}^2} = \sigma_{{\rm obs},\,i}^2+\sigma_{{\rm mod},\,i}^2$, $\sigma_{{\rm obs},\,i}$ is the observational uncertainty and $\sigma_{{\rm mod},\,i}$ accounts for the finite-photon Monte Carlo noise in the RT simulations. The sum is over unmasked spectral pixels within the fitting window, $F_i^{\rm obs}$ is the observed continuum-normalized flux, and $F_i^{\rm mod}$ is the forward-modeled RT spectrum. The priors are chosen to span the range of line-profile morphologies covered by the customized RT model library, with the parameter definitions summarized in Table~\ref{tab:model_parameters}. The physical and nuisance parameters are varied simultaneously during the fit.

The trained DNN is combined with a nested-sampling algorithm to identify the global best-fit solution and characterize the posterior distribution in the multi-dimensional parameter space. When necessary, we adaptively expand the explored parameter space through targeted resampling and retrain the emulator until the posterior is well contained within the adopted priors. We evaluate the emulator accuracy by comparing the DNN-predicted spectra with the original Monte Carlo RT spectra for validation models that are not used during training. As in \citet{Li2026a}, the residuals are dominated by minor local fluctuations rather than global mismatches in the line-profile morphology, indicating that the emulator provides a sufficiently accurate approximation to the RT calculations for parameter inference. We refer readers to Section~2.4 and Appendix~C of \citet{Li2026a} for further details.

\subsection{Best-fit Profiles and Morphological Classes}
\label{subsec:bestfit_morphology}

\begin{figure*}
\begin{center}
	\includegraphics[width=\textwidth]{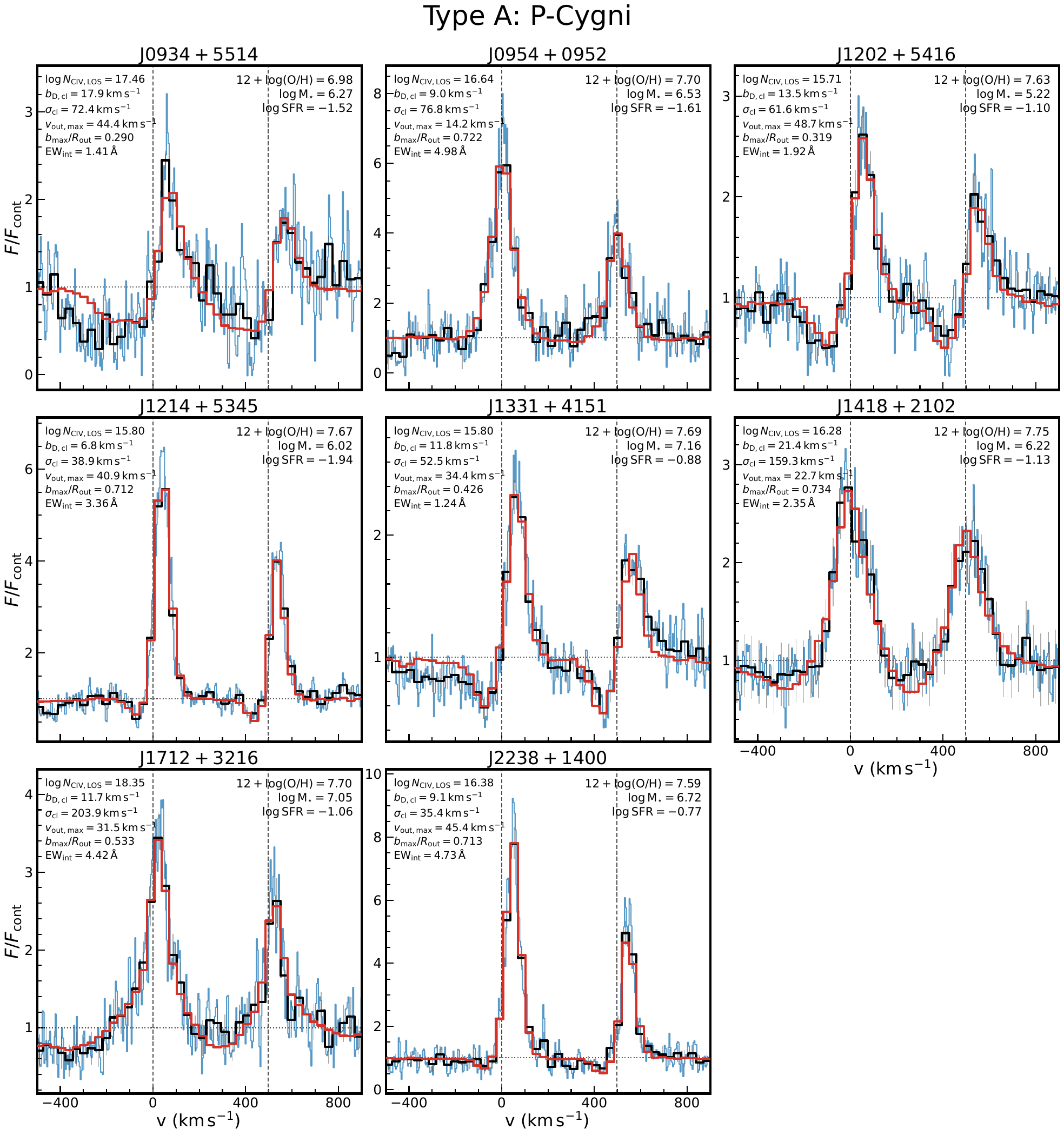}  
\end{center}
\caption{\textbf{Best-fit clumpy RT models for the objects in the P-Cygni profile class (Type~A).} In each panel, the blue curve shows the coadded, oversampled HST/COS \ion{C}{4} spectrum prior to rebinning, with a velocity sampling of $4\,{\rm km\,s^{-1}}$, while the black curve shows the rebinned spectrum with a velocity sampling of $32\,{\rm km\,s^{-1}}$. The red curve shows the best-fit RT model. The median best-fit RT model parameters are listed in the upper left, and the median host-galaxy properties are listed in the upper right. The vertical dotted lines mark the systemic velocities of the two components of the \ion{C}{4} doublet. These objects occupy a regime in which outflowing resonant absorption remains prominent despite partial emission infilling.}
\label{fig:type_A_CIV}
\end{figure*}

\begin{figure*}
\begin{center}
	\includegraphics[width=0.94\textwidth]{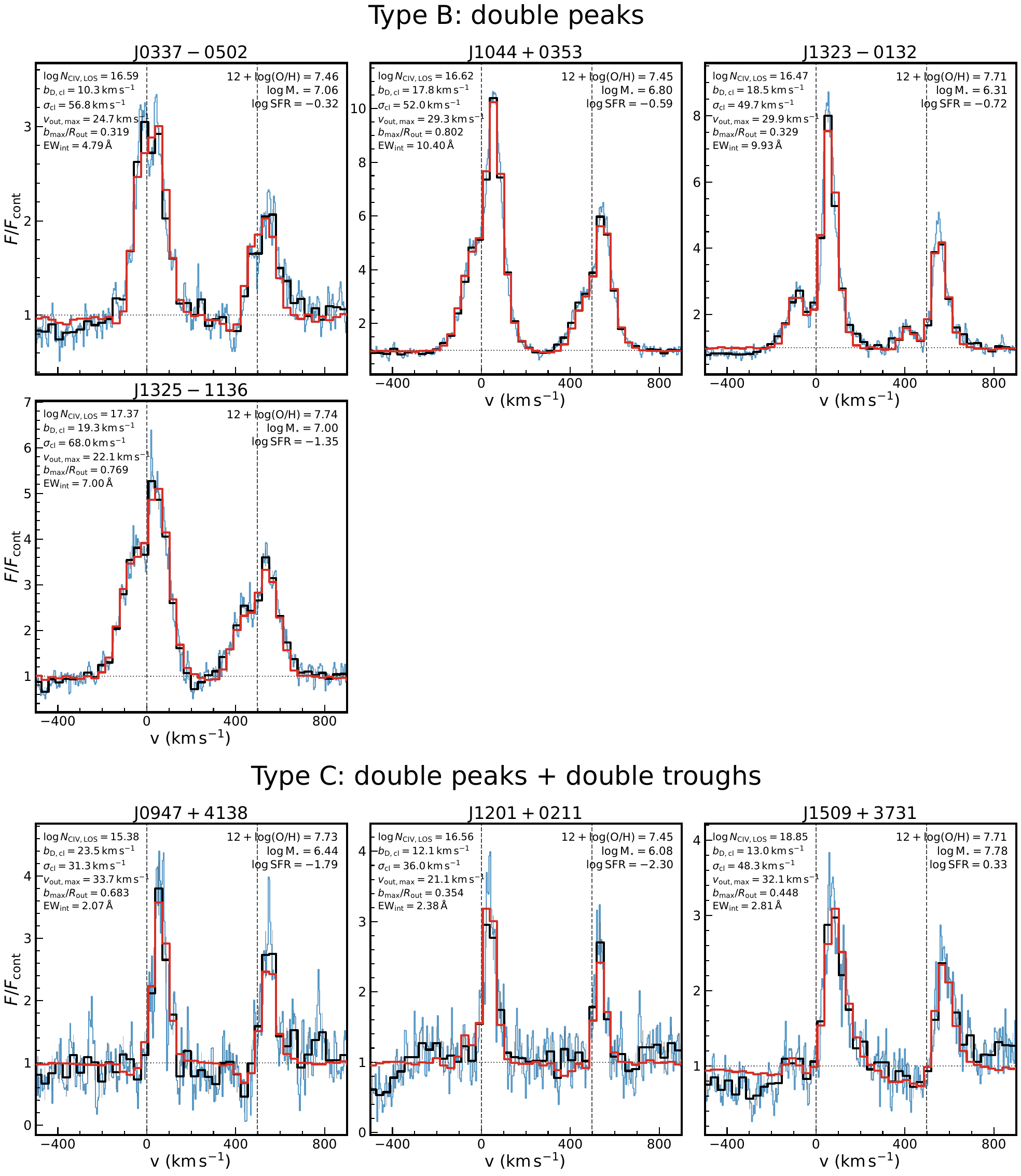}\\
\end{center}
\caption{\textbf{Best-fit clumpy RT models for the double-peaked emission (Type~B; upper rows) and double-peaked plus double-trough (Type~C; lower row) profile classes.} In each panel, the blue curve shows the coadded, oversampled HST/COS \ion{C}{4} spectrum prior to rebinning, with a velocity sampling of $4\,{\rm km\,s^{-1}}$, while the black curve shows the rebinned spectrum with a velocity sampling of $32\,{\rm km\,s^{-1}}$. The upper-row fits illustrate cases where intrinsic line emission and recovered scattered light strongly fill the blueshifted absorption, whereas the lower-row fits trace a narrower transition regime in which scattered emission partially fills in the absorption trough and imprints two troughs on the emergent profile.}
\label{fig:type_BC_CIV}
\end{figure*}

\begin{figure*}
\begin{center}
	\includegraphics[width=\textwidth]{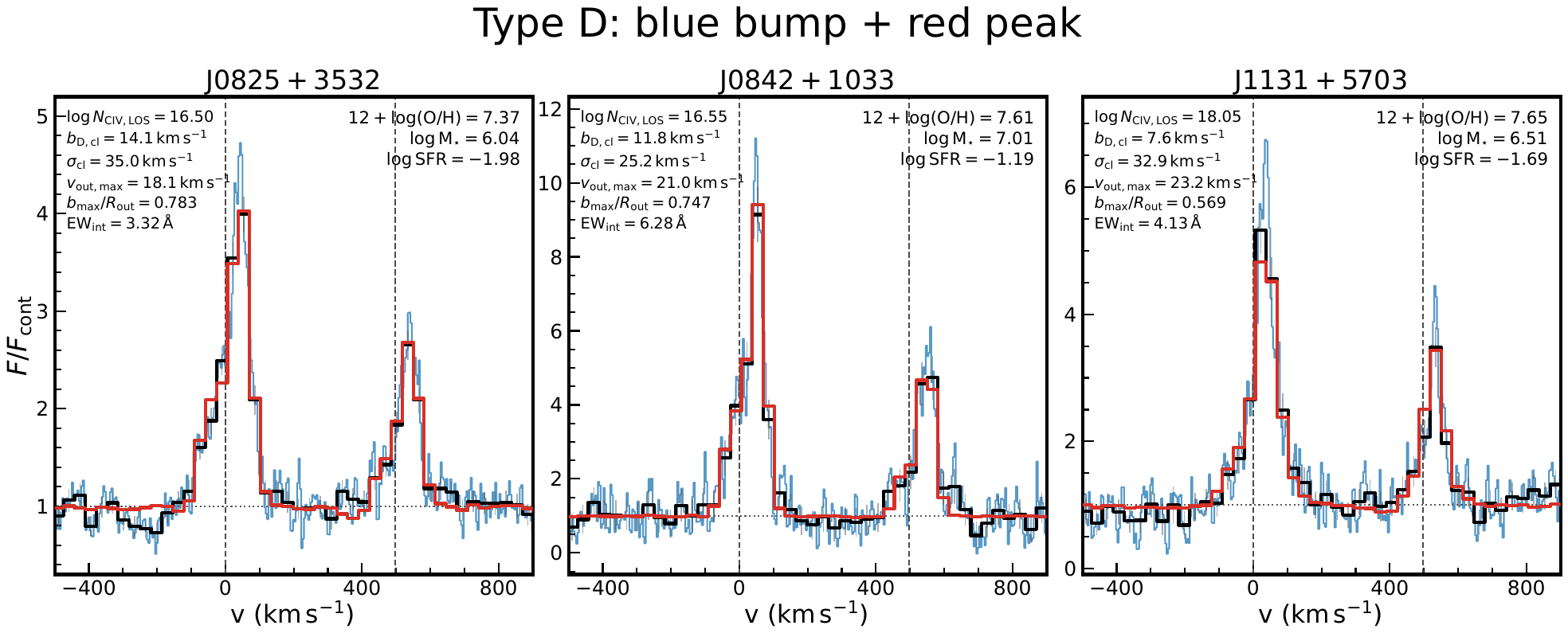}
\end{center}
\caption{\textbf{Best-fit clumpy RT models for the objects in the blue-bump plus red-peak profile class (Type~D).} In each panel, the blue curve shows the coadded, oversampled HST/COS \ion{C}{4} spectrum prior to rebinning, with a velocity sampling of $4\,{\rm km\,s^{-1}}$, while the black curve shows the rebinned spectrum with a velocity sampling of $32\,{\rm km\,s^{-1}}$. In these profiles, the blue-side emission appears as a subdominant bump that connects smoothly to a stronger red emission peak. These fits are consistent with efficient emission infilling of the blueshifted absorption by scattered photons, producing a subdominant blue bump rather than a distinct absorption trough or a prominent blue peak.}
\label{fig:type_D_CIV}
\end{figure*}

We present the best-fit \ion{C}{4}\,$\lambda\lambda1548,1550$ RT models in Figures~\ref{fig:type_A_CIV}--\ref{fig:type_D_CIV}. The best-fit model parameters are listed in the upper left of each panel (see also Table \ref{tbl:civ_rt_parameters} in the Appendix), while the host-galaxy properties are shown in the upper right. Overall, the models successfully reproduce the main line-profile morphologies of all 18 spectra. This agreement suggests that the observed diversity of \ion{C}{4} profiles can be largely explained within a single clumpy RT framework through variations in \ion{C}{4} column density, gas kinematics, intrinsic \ion{C}{4} equivalent width, and the aperture-dependent recovery of scattered emission.

Based on the observed and best-fit morphologies, the 18 profiles can be broadly described by the four profile classes introduced in Section~\ref{sec:composite_models}. We use these classes here only as descriptive rather than diagnostic categories, primarily to organize the best-fit results and highlight the main physical trends and degeneracies that underlie the observed profile diversity.

Figure~\ref{fig:type_A_CIV} shows the objects in the P-Cygni class (Type A) introduced in Section~\ref{sec:composite_models}. Their \ion{C}{4} profiles show blueshifted absorption together with redshifted emission, consistent with continuum photons being absorbed by outflowing C$^{3+}$ gas clumps along the line of sight and re-emitted through resonant scattering. This interpretation follows the composite RT experiments, where P-Cygni-like profiles arise when the emergent spectrum remains significantly shaped by continuum absorption. In the higher-outflow regime, scattered emission is shifted preferentially to the red side, while blue-side scattered emission is weakened and becomes less able to fill the blueshifted absorption trough.

This behavior is reflected in the best-fit models for the P-Cygni-class objects. As shown in Figure~\ref{fig:type_A_CIV}, five of the eight objects in this class have best-fit maximum outflow velocities close to or above $35\,{\rm km\,s^{-1}}$, placing them in the higher-outflow regime identified in the idealized experiments. These objects also show some of the clearest blueshifted absorption troughs in the sample. Their best-fit models therefore occupy a regime in which outflowing resonant absorption remains prominent. In addition, intrinsic \ion{C}{4} emission, aperture-dependent recovery of scattered light, and the fitted clump velocity parameters can all modulate the detailed profile shape. In particular, variations in $b_{\rm D,\,cl}$ and $\sigma_{\rm cl}$ change the widths of the absorption and emission features, while ${\rm EW}_{\rm int}$ and $b_{\rm max}/R_{\rm out}$ affect the degree of trough infilling and the strength of the red emission peak. In these objects, however, such effects do not fully erase the blueshifted absorption, leaving the P-Cygni-like morphology as the defining observational feature.

Figure~\ref{fig:type_BC_CIV} shows the objects in the double-peaked emission class (Type B) introduced in Section~\ref{sec:composite_models}. Compared to the Type~A profiles, the blueshifted absorption component is more strongly filled by scattered emission, so that it appears as a blue emission peak rather than an absorption trough, while the red emission peak remains prominent. This behavior is consistent with the composite RT experiments: in high-column-density, low-outflow-velocity models, increasing ${\rm EW}_{\rm int}$ can overwhelm the blueshifted continuum absorption and transform a P-Cygni-like profile into a double-peaked morphology.

The individual best-fit models illustrate different ways in which this double-peaked morphology can arise. J1323$-$0132 and J0337$-$0502 have relatively small best-fit effective apertures, resembling the high-column-density, low-outflow-velocity models with small $b_{\rm max}/R_{\rm out}$. In this regime, the recovery of extended scattered emission is limited, and the profile is shaped mainly by the balance between intrinsic \ion{C}{4} line emission and resonant absorption. By contrast, J1044+0353 and J1325$-$1136 are closer to the intermediate-aperture regime of the same model sequence. In these cases, photons scattered to larger impact parameters contribute more strongly to the observed spectrum, efficiently filling the absorption between the two peaks. The resulting profiles therefore remain double-peaked, but the trough separating the blue and red emission peaks is relatively shallow. The double-peaked emission fits therefore occupy a regime in which intrinsic \ion{C}{4} emission and aperture-dependent recovery of scattered light are strong enough to convert the blueshifted absorption characteristic of P-Cygni-like profiles into a blue emission peak or shoulder. At the same time, resonant scattering remains important, as indicated by the persistent peak separation and line-center absorption.

The lower panel of Figure~\ref{fig:type_BC_CIV} shows the objects in the double-peaked plus double-trough class (Type C) introduced in Section~\ref{sec:composite_models}. This morphology represents a particularly narrow transition regime between the P-Cygni-like profiles and the more strongly emission-filled double-peaked profiles. In the composite RT experiments, this behavior appears most clearly in the high-column-density, low-outflow-velocity models when the effective aperture is intermediate (corresponding to the red curve in the top row of Figure \ref{fig:composite_models}). In this regime, the intrinsic \ion{C}{4} emission is strong enough to lift part of the blueshifted absorption into emission, but not strong enough to erase the absorption structure entirely. The resulting profile contains both blue and red emission peaks while retaining one trough between the two peaks and an additional blueshifted trough.

Among the objects in this class, J1509+3731 provides the clearest example of the double-peak plus double-trough morphology. Its profile shows two well-defined emission peaks together with two distinct absorption troughs, closely resembling the narrow transition regime identified in the composite models. In addition, the red emission peak is significantly redshifted by $\sim100\,{\rm km\,s^{-1}}$, indicating stronger frequency diffusion and therefore leading to the largest inferred LOS \ion{C}{4} column density in our sample. The other objects in this class show the same qualitative morphology, although with lower contrast, partly because of their lower signal-to-noise ratios and narrower line profiles. These fits suggest that double-peak plus double-trough profiles occur when intrinsic \ion{C}{4} emission, resonant absorption, and aperture-dependent scattered-light recovery are finely balanced. In this regime, scattered photons are sufficient to produce a blue emission feature, while residual resonant absorption remains strong enough to imprint two troughs on the emergent profile.

Figure~\ref{fig:type_D_CIV} shows the objects in the blue-bump plus red-peak class (Type D) introduced in Section~\ref{sec:composite_models}, where the blue-side emission appears as a subdominant bump that connects smoothly to a dominant red emission peak. This morphology corresponds most closely to the high-column-density, low-outflow-velocity composite models with moderate to large ${\rm EW}_{\rm int}$ and relatively large effective apertures. In this regime, intrinsic \ion{C}{4} emission and spatially extended scattered photons efficiently fill the blueshifted absorption, while large $b_{\rm max}/R_{\rm out}$ allows blue-side scattered emission from large impact parameters to contribute strongly to the observed spectrum. The best-fit models for this class highlight the combined importance of intrinsic \ion{C}{4} emission and aperture-dependent recovery of scattered light.

Taken together, the four morphological classes form a continuous sequence rather than a set of discrete physical categories. This sequence reinforces the main result of the idealized RT experiments: the observed \ion{C}{4} profile diversity is not controlled by any single quantity, such as intrinsic line strength or outflow velocity alone, but by the coupled redistribution of line and continuum photons in both velocity space and projected radius. The velocity-resolved \ion{C}{4} morphology therefore encodes the joint effects of intrinsic \ion{C}{4} production, resonant scattering through C$^{3+}$ gas clumps, gas kinematics, and aperture-dependent recovery of spatially extended emission.

\section{UV-BPT Diagnostic Diagrams}\label{sec:uv_bpt}

The RT profile fitting presented above shows that the observed \ion{C}{4} line morphologies can be reproduced within a single clumpy RT framework. However, the profile shape alone does not uniquely determine the nature of the ionizing source that produces the intrinsic \ion{C}{4} photons. To assess whether the inferred \ion{C}{4} emission can be explained by stellar photoionization or instead requires a harder ionizing radiation field, we combine independently measured rest-frame UV emission-line ratios with the RT-inferred intrinsic \ion{C}{4} EWs and examine the locations of our galaxies on a set of UV diagnostic diagrams.

\begin{figure*}
\begin{center}
	\includegraphics[width=0.91\textwidth]{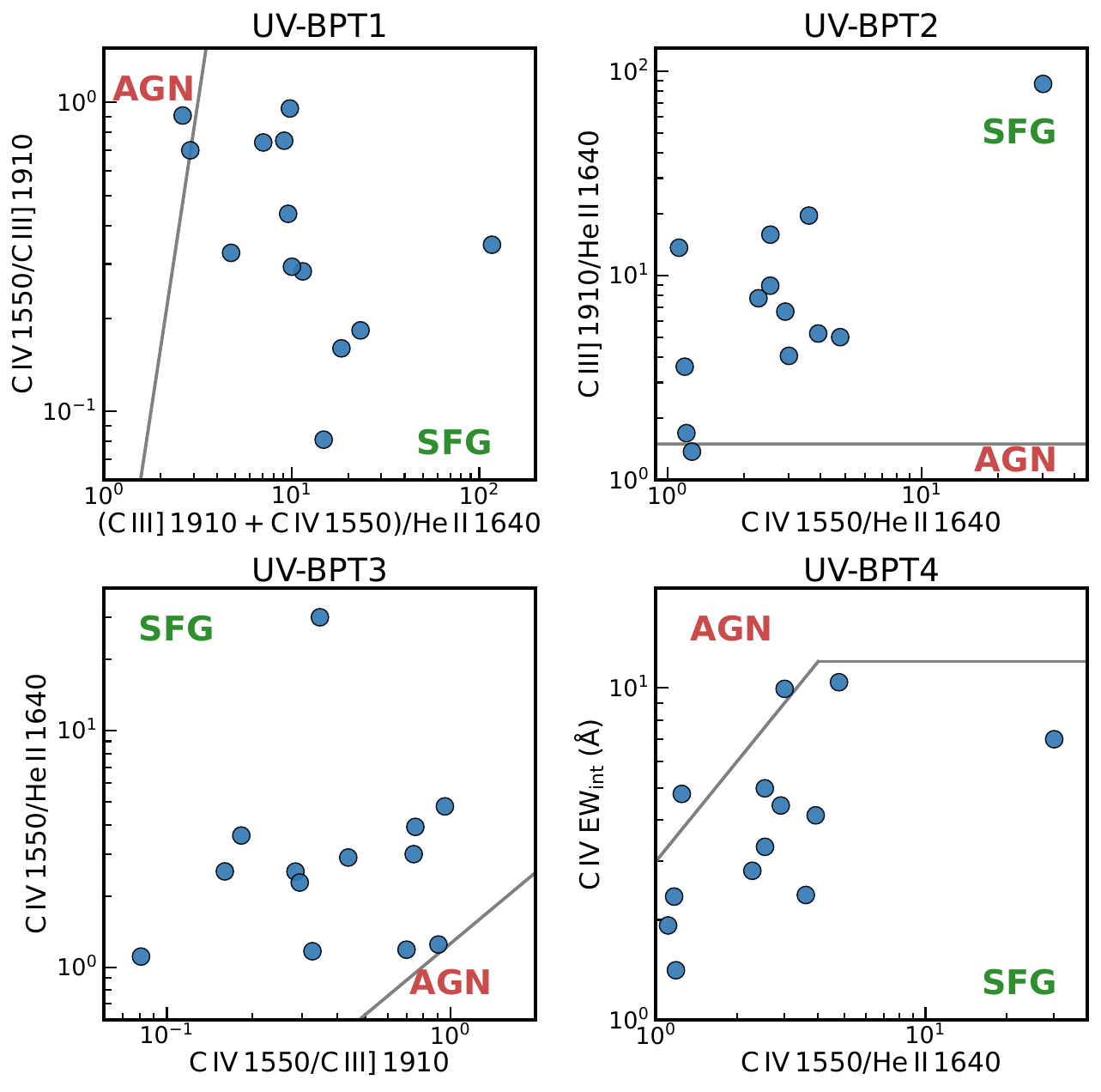}
\end{center}
\caption{\textbf{UV-BPT diagnostic diagrams for our sample, following the definitions of \citet{mascia23_civvandels}}. The four panels show different combinations of rest-frame UV line ratios involving \ion{C}{4} $\lambda\lambda1548,1550$, \ion{C}{3}] $\lambda\lambda1907,1909$, and \ion{He}{2} $\lambda1640$. Blue circles represent our sample, and the gray curves mark the empirical separation lines between the star-forming galaxy and AGN regimes. The first three panels use only observed UV line ratios, while UV-BPT4 combines the observed \ion{C}{4}/\ion{He}{2} ratio with the intrinsic \ion{C}{4} EW inferred from our RT modeling. Most sources occupy the star-forming galaxy regions, although a few objects with large inferred intrinsic \ion{C}{4} EWs lie close to the separation boundaries.}
\label{fig:uv_bpt_plot}
\end{figure*}

We measure the observed emission-line fluxes of several major rest-frame UV transitions used in these diagnostics, including \ion{C}{4} $\lambda\lambda1548,1550$, \ion{C}{3}] $\lambda\lambda1907,1909$, and \ion{He}{2} $\lambda1640$. Following \citet{mascia23_civvandels}, we place our galaxies on four UV-BPT diagrams, as shown in Figure~\ref{fig:uv_bpt_plot}. The first three diagrams are based entirely on observed UV line ratios, while the fourth diagram instead combines the observed \ion{C}{4}/\ion{He}{2} ratio with the intrinsic \ion{C}{4} EW inferred from the RT modeling. Most galaxies in our sample lie in the star-forming galaxy regions of the UV-BPT diagrams, suggesting that their observed UV line ratios are broadly consistent with stellar photoionization and do not require a dominant AGN contribution. Interestingly, three of the four double-peaked objects, J0337$-$0502, J1044+0353, and J1323$-$0132, lie either in the AGN regime or close to the AGN/SFG boundary in the EW-based UV-BPT4 diagram. Two of them, J1044+0353 and J1323$-$0132, have particularly large RT-inferred intrinsic \ion{C}{4} EWs, both $\gtrsim10\,{\rm \AA}$. These objects may therefore require a harder ionizing spectrum, more extreme ionization conditions, and/or a larger contribution from very young, metal-poor stellar populations. However, their locations in the other UV-BPT diagrams remain broadly consistent with the star-forming galaxy regime, so the UV diagnostics alone do not provide compelling evidence for AGN-dominated ionization.

\begin{figure}
\begin{center}
	\includegraphics[width=\columnwidth]{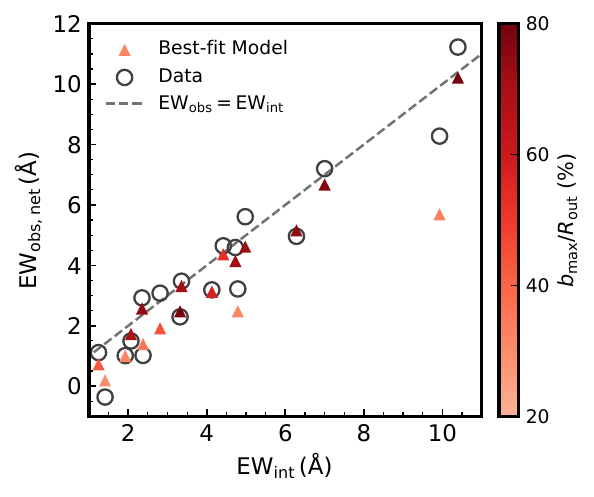}
\end{center}
\caption{\textbf{Comparison between intrinsic and observed net \ion{C}{4} equivalent widths.} The x-axis shows the intrinsic \ion{C}{4} EW inferred from the best-fit RT model for each galaxy. Hollow black circles show the observed net \ion{C}{4} EW measured directly from the data, while filled triangles show the net \ion{C}{4} EW measured from the corresponding best-fit model. The triangle colors indicate the best-fit maximum normalized aperture, $b_{\max}/R_{\rm out}$. The dashed line marks ${\rm EW}_{\rm obs,\,net}={\rm EW}_{\rm int}$. Most best-fit models lie below the one-to-one relation, especially for smaller $b_{\max}/R_{\rm out}$, illustrating that resonant scattering and aperture losses generally make the observed net EW smaller than the intrinsic EW required for the source.}
\label{fig:ew_comparison}
\end{figure}

A complementary check comes from comparing the intrinsic \ion{C}{4} EW required by the RT models with the net \ion{C}{4} EW measured from the emergent spectra. Figure~\ref{fig:ew_comparison} shows this comparison for both the observed data, shown as hollow black circles, and the best-fit RT models, shown as filled triangles color-coded by the best-fit $b_{\max}/R_{\rm out}$. Consistent with the discussion in Section~\ref{sec:ewint_vs_ewobs}, the best-fit models generally lie below the one-to-one relation. Moreover, models with smaller inferred $b_{\max}/R_{\rm out}$ tend to show larger deficits in ${\rm EW}_{\rm obs,\,net}$ relative to ${\rm EW}_{\rm int}$, because smaller apertures recover a smaller fraction of the spatially extended scattered emission. The observed net \ion{C}{4} EWs measured directly from the data show a similar behavior: ten objects have ${\rm EW}_{\rm obs,\,net}<{\rm EW}_{\rm int}$, and eight lie approximately on the ${\rm EW}_{\rm obs,\,net}={\rm EW}_{\rm int}$ relation. Overall, both the model and data measurements support the interpretation that ${\rm EW}_{\rm obs,\,net}$ is generally a lower limit on the intrinsic \ion{C}{4} EW required at the source.

Taken together, the UV-BPT diagrams and EW comparison support a picture in which strong \ion{C}{4} emission in our sample is primarily associated with compact, highly ionized star-forming regions rather than AGN-dominated sources. At the same time, combining UV line ratios with RT-inferred intrinsic \ion{C}{4} EWs identifies a small subset of objects with more extreme ionizing conditions. This comparison highlights the value of combining resonant-line RT modeling with non-resonant UV diagnostics: the RT modeling separates intrinsic \ion{C}{4} production from resonant scattering and aperture effects, while the UV-BPT diagrams provide an independent constraint on the hardness of the ionizing radiation field.

\section{Discussion}
\label{sec:discussion}

\subsection{From Emission-dominated to Absorption-dominated \ion{C}{4} Profiles}
\label{subsec:emission_vs_absorption}

\begin{figure*}
\begin{center}
	\includegraphics[width=\textwidth]{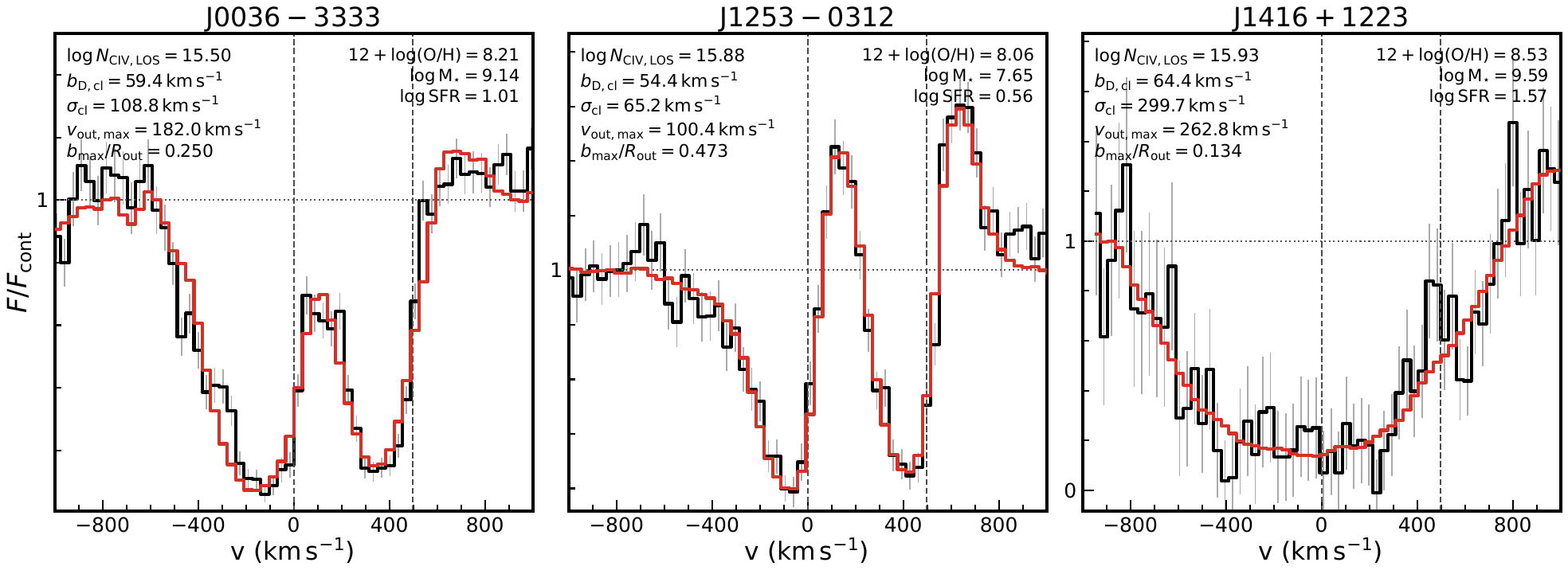}
\end{center}
\caption{\textbf{Representative absorption-dominated \ion{C}{4} profiles from the CLASSY sample.} Black and red curves show the observed spectra and best-fit RT models, respectively, with gray error bars indicating the observational uncertainties. These examples span distinct double-trough profiles to a broad, blended absorption trough, illustrating morphologies that contrast with the emission-selected sample analyzed in this work.}
\label{fig:classy_example_profiles}
\end{figure*}

The 18 galaxies modeled in this work were selected to have clearly detected \ion{C}{4} emission, and their profiles therefore occupy a relatively narrow, emission-dominated subset of the broader diversity of \ion{C}{4} profiles. To place these objects in a wider context, Figure~\ref{fig:classy_example_profiles} shows three representative \ion{C}{4} profiles from the CLASSY sample. Compared with the galaxies analyzed in this work, the CLASSY sample contains a much larger fraction of broad, absorption-dominated \ion{C}{4} profiles\footnote{Five CLASSY galaxies overlap with the sample analyzed in this work: J0337$-$0502, J0934+5514, J1044+0353, J1323$-$0132, and J1418+2102.}. J0036$-$3333 exhibits two broad absorption troughs, each blueshifted by more than $100\,{\rm km\,s^{-1}}$ relative to its corresponding doublet transition. J1253$-$0312 shows a similar double-trough morphology but more pronounced re-emission peaks. Within our RT framework, its larger inferred effective aperture allows a greater fraction of resonantly scattered photons to be recovered. J1416+1223 represents an even broader case, in which the absorption associated with the two doublet components is blended into a single wide absorption trough. Together, these examples illustrate profile morphologies that are qualitatively different from the narrower, emission-dominated spectra of the sample modeled in this work.

This difference is also reflected in the best-fit RT parameters. The CLASSY profiles generally require larger $b_{\rm D,\,cl}$, $\sigma_{\rm cl}$, and $v_{\rm out,\,max}$ than the 18 emission-dominated galaxies analyzed in this work. These larger clump Doppler parameters, macroscopic velocity dispersions, and outflow velocities indicate that the C$^{3+}$ gas clumps in the CLASSY galaxies are more dynamically disturbed and associated with stronger outflows. Most of these absorption-dominated profiles also have ${\rm EW}_{\rm net}<0$, meaning that they can be reproduced without invoking intrinsic \ion{C}{4} line emission; continuum photons undergoing resonant absorption and scattering are sufficient to reproduce the observed profiles. By contrast, the emission-selected galaxies in this work typically occupy regimes in which intrinsic \ion{C}{4} emission and aperture-dependent recovery of scattered light play a more significant role in shaping the emergent profile.

This systematic difference becomes more understandable in the context of the host-galaxy properties. Compared with the 18 galaxies modeled in this work, the CLASSY galaxies extend to higher stellar masses and higher SFRs. Their broader \ion{C}{4} profiles may therefore reflect differences in both the gravitational environment and stellar feedback, which can influence the bulk velocities and velocity dispersions of the C$^{3+}$ gas clumps. This suggests that \ion{C}{4} profile morphology is shaped not only by the ionization conditions and intrinsic line production, but also by the global dynamical state of the gas surrounding the galaxy.

\begin{figure*}
\begin{center}
	\includegraphics[width=0.82\textwidth]{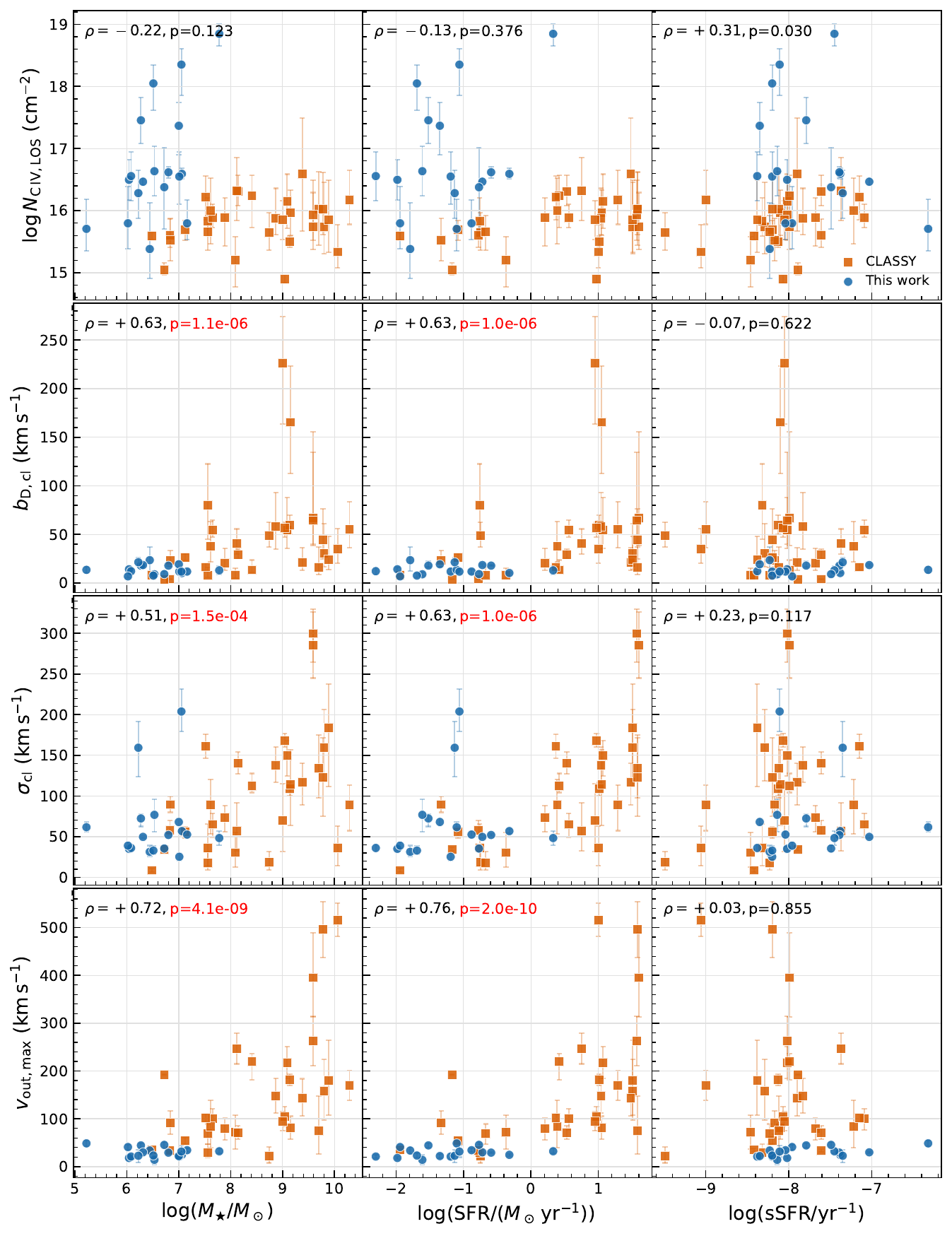}
\end{center}
\caption{\textbf{Relations between \ion{C}{4} RT parameters and host-galaxy properties.} Blue circles represent the 18 emission-selected galaxies analyzed in this work, and orange squares represent non-overlapping CLASSY galaxies. From left to right, the columns show stellar mass, SFR, and sSFR. From top to bottom, the rows show the fitted LOS \ion{C}{4} column density, clump Doppler parameter, macroscopic clump velocity dispersion, and maximum clump outflow velocity. Points and error bars indicate posterior medians and the 16th--84th percentile intervals. Each panel lists the Spearman correlation coefficient, $\rho$, and its two-sided $p$-value for the combined sample, calculated using the posterior medians. $p$-values indicating correlations significant at $>3\sigma$ are highlighted in red. The three kinematic parameters correlate positively with stellar mass and SFR, but show no significant association with sSFR.}
\label{fig:compare_with_classy}
\end{figure*}

We examine these connections more quantitatively in Figure~\ref{fig:compare_with_classy}, comparing four inferred \ion{C}{4} RT parameters with stellar mass, $M_\star$, SFR, and sSFR for the 18 emission-selected galaxies and non-overlapping CLASSY galaxies. The emission-selected sample alone occupies a relatively restricted range of $M_\star$, SFR, and RT parameters, making it difficult to identify robust scaling relations. Once combined with the CLASSY sample, however, the parameter space expands substantially, and several correlations become apparent. The strongest trends ($>3\sigma$ significant) involve the kinematic parameters: $b_{\rm D,\,cl}$, $\sigma_{\rm cl}$, and $v_{\rm out,\,max}$ all increase with increasing $M_\star$ and SFR. None of these three parameters shows a significant correlation with sSFR, suggesting that the inferred gas kinematics are more closely associated with absolute stellar mass and SFR than with star formation per unit stellar mass. By contrast, the LOS \ion{C}{4} column density shows no statistically compelling correlation with stellar mass or SFR. Thus, the transition from narrow, emission-dominated profiles to broad, absorption-dominated CLASSY profiles appears to reflect changes in gas kinematics and the relative strength of intrinsic \ion{C}{4} emission, rather than simply differences in the column density of C$^{3+}$ gas clumps.

These comparisons reinforce one of the central conclusions of this work: velocity-resolved \ion{C}{4} profiles provide information that cannot be recovered from integrated line strength alone. \ion{C}{4} profile morphology therefore traces both intrinsic \ion{C}{4} production and the dynamical state of the C$^{3+}$-bearing scattering gas. Their correlations with stellar mass and SFR further suggest that \ion{C}{4} RT modeling can connect rest-UV spectral morphology to feedback-driven gas dynamics in compact star-forming galaxies and their more massive CLASSY counterparts.

\subsection{Implications for Reionization-Era \ion{C}{4} Emitters}
\label{subsec:highz_implications}

Strong nebular \ion{C}{4} emission is increasingly observed in galaxies at $z\gtrsim6$, where it is commonly interpreted as evidence for unusually hard ionizing radiation fields and highly ionized gas \citep[e.g.,][]{Stark2015,Topping2025}. Our RT results add an important dimension to this interpretation: the observed strength and morphology of \ion{C}{4} depend not only on the intrinsic production of C$^{3+}$ and \ion{C}{4} photons, but also on their resonant transfer through the surrounding high-ionization gas. RT modeling therefore provides a way to separate these effects and constrain the intrinsic \ion{C}{4} equivalent width, ${\rm EW}_{\rm int}$, which more directly reflects the intrinsic nebular line production than the observed equivalent width alone. Without accounting for resonant scattering and aperture-dependent redistribution, the observed \ion{C}{4} strength cannot be uniquely mapped onto intrinsic line production. Galaxies with comparable intrinsic \ion{C}{4} emission can therefore exhibit substantially different observed profiles and EWs depending on their C$^{3+}$ column densities, gas kinematics, and the spatial redistribution of scattered photons.

This separation between intrinsic \ion{C}{4} production and RT effects may be particularly important for understanding ionizing-photon escape during reionization. Most empirical constraints on LyC escape probe neutral gas or the ionizing continuum near the Lyman limit \citep[see e.g.,][]{Jaskot2025}, whereas nebular \ion{C}{4} requires photons energetic enough to produce C$^{3+}$ ($\gtrsim48$ eV) \citep[e.g.,][]{Stark2015,Berg2019}. Because \ion{C}{4} is resonant, its emergent profile also carries information about the opacity and kinematics of the C$^{3+}$-bearing gas through which the photons propagate. The escape channels probed by \ion{C}{4} may therefore differ from those traced by neutral hydrogen, as previously suggested by \citet{Berg2019}, providing a potential window onto the highly ionized phase of the ISM and CGM that is largely inaccessible to H\,{\sc i}-based diagnostics alone.

Our local \ion{C}{4} emitters therefore provide a laboratory for investigating how the conditions governing photon escape vary with ionization state and, potentially, with photon energy. Our analysis provides a first step toward constraining the escape of harder ionizing radiation by probing the C$^{3+}$-bearing gas through which such radiation propagates. The diversity of \ion{C}{4} profiles and their sensitivity to C$^{3+}$ column density, gas kinematics, and resonant transfer demonstrate that the transparency of this highly ionized phase cannot be inferred from intrinsic \ion{C}{4} production alone. Quantifying the contribution of these harder photons to reionization will ultimately require direct or indirect constraints on their escape fraction. Joint modeling of velocity-resolved \ion{C}{4} with Ly$\alpha$, lower-ionization resonant lines, and direct LyC measurements available for low-redshift analogs \citep[e.g.,][]{Schaerer2022,Jaskot2025,Li2026a, Li2026b} will be an important next step toward constraining the energy dependence of the escaping ionizing spectrum and assessing the role of hard ionizing photons in reionization.

\subsection{Caveats and Future Prospects}
\label{subsec:caveats}

Several caveats should be considered when interpreting the inferred model parameters. First, our RT framework assumes an idealized clumpy geometry and a simplified description of gas kinematics. The fitted parameters are therefore best regarded as effective quantities that reproduce the observed \ion{C}{4} profiles, rather than unique measurements of the underlying three-dimensional gas distribution. In particular, $b_{\rm max}/R_{\rm out}$ is an effective aperture parameter that captures the fraction of scattered emission recovered in the observed spectrum. Without spatially resolved constraints, it should not be translated directly into a physical size for the \ion{C}{4}-emitting halo.

Second, the emergent \ion{C}{4} spectrum depends on both intrinsic line production and the distribution and physical state of the scattering gas clumps. Our models separate intrinsic \ion{C}{4} production from resonant scattering phenomenologically through ${\rm EW}_{\rm int}$ and the RT parameters, but they do not self-consistently predict the ionization structure of the gas. Combining \ion{C}{4} RT modeling with photoionization calculations will be necessary to connect the inferred \ion{C}{4} column densities and intrinsic equivalent widths to metallicity, ionization parameter, stellar population age, and LyC escape fraction.

Future observations can help resolve these uncertainties. Spatially resolved rest-UV spectroscopy would help determine how much \ion{C}{4} emission escapes at large projected radii and would provide a more physical calibration of the effective aperture parameter. Larger samples with velocity-resolved \ion{C}{4}, C\,{\sc iii]}, He\,{\sc ii}, Ly$\alpha$, and Mg\,{\sc ii} will also make it possible to connect the C$^{3+}$-bearing gas traced by \ion{C}{4} to the neutral and low-ionization gas phases that regulate LyC escape. Ultimately, combining \ion{C}{4} RT modeling with photoionization calculations and multi-line resonant RT modeling would help break degeneracies between intrinsic line production and scattering, linking \ion{C}{4} profile morphology more directly to multiphase gas structure and the conditions that permit LyC escape in star-forming galaxies.

\section{Conclusions}
\label{sec:conclusions}

We have presented a systematic radiative transfer analysis of velocity-resolved \ion{C}{4}\,$\lambda\lambda1548,1550$ profiles in 18 local star-forming galaxies with detected \ion{C}{4} emission. Using the clumpy RT framework \texttt{PEACOCK}, we modeled \ion{C}{4} as a resonant doublet shaped by both intrinsic emission and the resonant redistribution of line and continuum photons through C$^{3+}$ gas clumps. Our main conclusions are as follows.

\begin{enumerate}

\item \textbf{A single clumpy RT framework reproduces the observed diversity of \ion{C}{4} profiles.} The models reproduce the main morphologies of all 18 galaxies: P-Cygni-like, double-peaked, double-peak plus double-trough, and blue-bump plus red-peak profiles. The four observed morphological classes arise from the interplay of column density, gas kinematics, intrinsic line strength, and aperture-dependent recovery of scattered emission, rather than uniquely identifying distinct physical configurations. In particular, emission infilling can transform a blueshifted absorption trough into
a blue emission peak or bump, connecting apparently different
profile morphologies.

\item \textbf{Emission infilling can bias outflow velocities inferred from absorption troughs.} Recovering more scattered emission can both weaken an absorption trough and displace its apparent minimum. In one illustrative model with $v_{\rm out,\,max}=20\,\mathrm{km\,s^{-1}}$, infilling shifts
the absorption minimum to $\sim-50\,\mathrm{km\,s^{-1}}$. The observed trough can therefore suggest a substantially faster
outflow than the model's actual bulk motion. Reliable kinematic
inferences require jointly modeling the transfer of intrinsic line
and continuum photons, including aperture-dependent emission
recovery, rather than equating the trough velocity with the
outflow speed.

\item \textbf{Observed net equivalent width is not a unique measure
of intrinsic \ion{C}{4} production.}
Resonant scattering redistributes photons in both frequency and
projected radius, allowing finite apertures to miss a substantial
fraction of the emergent emission. In our dust-free experiments,
the intrinsic EW is recovered when all escaping photons are
collected, whereas smaller apertures can substantially suppress
the observed net EW. The fitted galaxies show a similar tendency,
with larger EW deficits generally associated with smaller effective
apertures. Weak net emission, or even net absorption, therefore
need not imply weak intrinsic \ion{C}{4} production. Interpreting
\ion{C}{4} strength as a diagnostic of the ionizing source requires
accounting for scattering and aperture losses.

\item \textbf{The UV diagnostics broadly support stellar photoionization.} Most galaxies have UV line ratios consistent with stellar ionizing sources, without requiring an AGN-dominated contribution. J1044+0353 and J1323$-$0132 stand out with RT-inferred intrinsic \ion{C}{4} EWs of approximately $10\,\mathrm{\AA}$ and lie near the
EW-based diagnostic boundary. However, proximity to this boundary is not itself evidence for an AGN, particularly when intrinsic EWs
are combined with observed, transfer-affected line ratios.
Self-consistent photoionization and RT calculations are needed
to distinguish the effects of the ionizing source from those of
the scattering gas.

\item \textbf{Gas kinematics provide the clearest connection between \ion{C}{4} profiles and host-galaxy properties.} Relative to our emission-selected galaxies, the broader CLASSY sample includes more absorption-dominated profiles with larger $v_{\rm out,\,max}$, $b_{\rm D,\,cl}$, and $\sigma_{\rm cl}$.
Across the combined sample, these parameters correlate positively
with stellar mass and SFR, but not significantly with sSFR; the
LOS \ion{C}{4} column density shows no comparably significant
correlation with stellar mass or SFR. The morphological contrast is
thus more closely associated with gas kinematics and intrinsic
emission strength than with column density alone.

\end{enumerate}

Overall, our results demonstrate that \ion{C}{4} is not only an empirical tracer of hard ionizing radiation, but also a resonant-line probe of the highly ionized gas through which \ion{C}{4} photons propagate. Its velocity-resolved morphology records how intrinsic \ion{C}{4} line photons and continuum photons are redistributed by scattering in C$^{3+}$-bearing gas, both in velocity space and projected radius. As a result, \ion{C}{4} profiles encode valuable information about gas column density, outflow kinematics, velocity dispersion, and aperture-dependent scattered-light recovery that cannot be obtained from integrated fluxes, equivalent widths, or line ratios alone.

Velocity-resolved \ion{C}{4} RT modeling therefore provides a physically motivated way to separate intrinsic \ion{C}{4} production from RT effects. Combined with rest-UV line-ratio diagnostics, it can help connect the ionizing spectra of compact star-forming galaxies to the structure and kinematics of their C$^{3+}$-bearing gas, offering insight into the interplay between ionizing sources and stellar feedback. This framework will be particularly valuable for interpreting \ion{C}{4} emission in local reionization analogs and in high-redshift galaxies approaching the epoch of reionization, where \ion{C}{4} is increasingly used as a probe of hard radiation fields and possible escape of ionizing photons.

\begin{acknowledgements}

This work is based on observations obtained with the NASA/ESA
\textit{Hubble Space Telescope} and retrieved from the Mikulski
Archive for Space Telescopes (MAST) at the Space Telescope Science
Institute (STScI). STScI is operated by the Association of
Universities for Research in Astronomy, Inc., under NASA contract
NAS5-26555. The observations are associated with HST programs
11523, 11579, 13312, 13788, 14080, 14628, 15193, 15465, 15646,
15840, 15881, 16071, 16227, 16294, 16445, 16643, and 16677.
Support for program GO-16643 was provided by NASA through a
grant from STScI.
The HST/COS observations listed in Table~\ref{tbl2:observations} are available at \dataset[doi:10.17909/m2j2-3057]{https://doi.org/10.17909/m2j2-3057}.

We thank the members of the COS Legacy Archive Spectroscopic
SurveY (CLASSY) collaboration for their contributions to the
observations and data products that enabled this work.
This study makes use of CLASSY high-level science products
available through MAST at
\dataset[doi:10.17909/m3fq-jj25]{https://doi.org/10.17909/m3fq-jj25}.

The computations were carried out using the Advanced Research
Computing at Hopkins (ARCH) core facility
(\url{https://rockfish.jhu.edu}), supported by the National
Science Foundation under grant OAC-1920103.

MG acknowledges support from the Max Planck Society through
the Max Planck Research Group and from the European Union
through ERC-2024-STG 101165038 (ReMMU).

Funding for the Sloan Digital Sky Survey (SDSS) and SDSS-II
has been provided by the Alfred P. Sloan Foundation, the
Participating Institutions, the National Science Foundation,
the U.S. Department of Energy, the National Aeronautics and
Space Administration, the Japanese Monbukagakusho, the Max
Planck Society, and the Higher Education Funding Council for
England. The SDSS website is \url{http://www.sdss.org/}.
The SDSS is managed by the Astrophysical Research Consortium
(ARC) for the Participating Institutions. The Participating
Institutions are the American Museum of Natural History,
Astrophysical Institute Potsdam, University of Basel,
University of Cambridge, Case Western Reserve University,
The University of Chicago, Drexel University, Fermilab,
the Institute for Advanced Study, the Japan Participation
Group, The Johns Hopkins University, the Joint Institute
for Nuclear Astrophysics, the Kavli Institute for Particle
Astrophysics and Cosmology, the Korean Scientist Group,
the Chinese Academy of Sciences (LAMOST), Los Alamos
National Laboratory, the Max-Planck-Institute for Astronomy
(MPIA), the Max-Planck-Institute for Astrophysics (MPA),
New Mexico State University, Ohio State University,
University of Pittsburgh, University of Portsmouth,
Princeton University, the United States Naval Observatory,
and the University of Washington.

OpenAI Codex (OpenAI; \url{https://openai.com/codex/}) was used to assist with language editing and manuscript formatting.

\end{acknowledgements}

\appendix
\section{HST/COS Observations}\label{sec:observations}
Table~\ref{tbl2:observations} summarizes the new and archival
\textit{HST}/COS observations of the 18 galaxies in our
\ion{C}{4} emission sample. The observations are grouped by
target and observing program, with the dataset identifiers,
instrumental configurations, and exposure times provided to
document the spectroscopic coverage. The velocity-resolved
\ion{C}{4} profile analysis uses the medium-resolution FUV
spectra. The table also includes archival G140L and NUV G185M
observations, documenting the broader UV spectroscopic
coverage of the sample.
\startlongtable
\begin{deluxetable*}{rcccccc}
\tabletypesize{\footnotesize}
\tablewidth{\textwidth}
\tablecaption{HST/COS observations of the \ion{C}{4} emission sample\label{tbl2:observations}}
\tablehead{
\colhead{Target} &
\colhead{PID} &
\colhead{PI} &
\colhead{Dataset} &
\colhead{Grating} &
\colhead{$\lambda_{\rm cen}$\,(\AA)} &
\colhead{$t_{\rm exp}$\,(s)}
}
\startdata
1. J0337$-$0502 & 15193 & Aloisi    & LDN759010 & G130M & 1222 & 4581  \\
                &       &           & LDN709020 & G160M & 1623 & 5605  \\
                &       &           & LDN759020 & G160M & 1623 & 5345  \\[1.5ex]
                & 13788 & Wofford   & LCNE03010 & G160M & 1611 & 4959  \\
                &       &           & LCNE03020 & G185M & 1835 & 8512  \\
\tableline
2. J0825$+$3532 & 16643 & Gazagnes  & LEQR13020 & G130M & 1291 & 5486  \\
                &       &           & LEQR13010 & G160M & 1611 & 2141  \\
                &       &           & LEQR12010 & G160M & 1611 & 7629  \\[1.5ex]
               	& 15881 & Stark     & LE4Z02010 & G160M & 1533 & 5234  \\
                &       &           & LE4Z53010 & G160M & 1533 & 2400  \\
                &       &           & LE4Z01010 & G160M & 1533 & 5234  \\
\tableline
3. J0842$+$1033 & 16677 & Stark     & LENE02010 & G160M & 1533 & 7559  \\
                &       &           & LENE51010 & G160M & 1533 & 2173  \\
\tableline
4. J0934$+$5514 & 15193 & Aloisi    & LDN707010 & G130M & 1222 & 2352  \\[1.5ex]
                & 11579 & Aloisi    & LB7H71010 & G130M & 1291 & 7762  \\
                &       &           & LB7H72010 & G130M & 1291 & 7762  \\[1.5ex]
                & 11523 & Green     & LB6L1A030 & G130M & 1291 & 3409  \\ 
                &       &           & LB6L2A0C0 & G160M & 1589 & 1768  \\
                &       &           & LB6L2A0D0 & G160M & 1600 & 1530  \\
                &       &           & LB6L2A0E0 & G160M & 1611 & 2140  \\
                &       &           & LB6L2A0F0 & G160M & 1623 & 1872  \\
                &       &           & LB6L2A0G0 & G185M & 1900 & 2098  \\
                &       &           & LB6L2A0H0 & G185M & 1913 & 4049  \\
                &       &           & LB6L2A0I0 & G185M & 1921 & 1949  \\
\tableline
5. J0947$+$4138 & 16643 & Gazagnes  & LEQRZ0010 & G130M & 1291 & 7568  \\
                &       &           & LEQR09020 & G130M & 1291 & 1270  \\
                &       &           & LEQRZ8010 & G160M & 1533 & 2779  \\
               	&       &           & LEQR09010 & G160M & 1533 & 6004  \\
                &       &           & LEQR14010 & G160M & 1533 & 1952  \\
\tableline
6. J0954$+$0952 & 16677 & Stark     & LENE03010 & G160M & 1533 & 7551  \\
                &       &           & LENE52010 & G160M & 1533 & 2173  \\
\tableline
7. J1044$+$0353 & 15840 & Berg      & LE2438010 & G130M & 1291 & 6854  \\[1.5ex]
                & 15465 & Berg      & LDSZ01010 & G160M & 1589 & 6439  \\[1.5ex]
               	& 15646 & Stark     & LDXT07010 & G160M & 1533 & 13183 \\
                &       &           & LDXT08010 & G160M & 1533 & 13183 \\
\tableline
8. J1131$+$5703	& 14628 & Berg	    & LDAF07010	& G140L	& 1280 & 2192  \\[1.5ex] 	
		        & 16677 & Stark	    & LENE05010	& G160M & 1533 & 8309  \\	
				&       &           & LENE06010 &       & 1533 & 8309  \\
\tableline
9. J1201$+$0211	& 13312 & Berg	    & LCAW04010 & G140L	& 1280 & 2056  \\[1.5ex]	
		        & 16677 & Stark	    & LENE07010 & G160M	& 1533 & 7608  \\	
				&      	&           & LENE08010	&       & 1533 & 7608  \\	
\tableline
10. J1202$+$5416& 16227 & Bowen	    & LEEK03010 & G130M	& 1291 & 7640 \\	
				&       &           & LEEK04010	&       &      & 7640  \\[1.5ex]
		        & 15881 & Stark	    & LE4Z04010 & G160M	& 1533 & 5488  \\	
				&       &           & LE4Z05010	&       &      & 5488  \\	
				&       &           & LE4Z06010	&       &      & 2521  \\	
\tableline
11. J1214$+$5345& 14628 & Berg	    & LDAF04010 & G140L	& 1280 & 2228  \\	
		        & 16677 & Stark	    & LENE09010 & G160M	& 1533 & 8225  \\
				&       &           & LENE10010 &       & 	   & 8225  \\
\tableline
12. J1323$-$0132& 15840 & Berg	    & LE2444010 & G130M & 1291 & 6914  \\	
			    &       & 		    & LE2445010	& G160M & 1623 & 11391 \\[1.5ex]
		        & 16677 & Stark	    & LENE11010 & G160M	& 1533 & 7636  \\	
				&       &           & LENE12010	&       &      & 7636  \\	
\tableline
13. J1325$-$1136& 16445 & Kumari	& LEID01010 & G130M	& 1291 & 5123  \\[1.5ex]	
		        & 16071 & Kumari	& LE9701010 & G160M	& 1623 & 4528  \\[1.5ex]
		        & 16294 & Eggen	    & LEDT04010 & G160M	& 1533 & 10456 \\[1.5ex]	
		        & 16071 & Kumari	& LE9702010 & G185M	& 1913 & 13429 \\	
\tableline
14. J1331$+$4151& 16643 & Gazagnes	& LEQR06020 & G130M	& 1291 & 2656  \\	
			    &       &           & LEQR07010 & G160M	& 1533 & 3366  \\	
				&       &           & LEQR06010	&       &      & 2330  \\[1.5ex]
				& 16677 & Stark     & LENE53010	& G160M & 1533 & 5204  \\	
				&       &           & LENE54010	&       &      & 5204  \\	
\tableline
15. J1418$+$2102& 15840 & Berg	    & LE2435010 & G130M	& 1291 & 9590  \\	
				&       &           & LE2437010	&       &      & 6860  \\[1.5ex]	
		        & 15465 & Berg	    & LDSZ02010 & G160M	& 1589 & 12374  \\	
\tableline
16. J1509$+$3731& 14080 & Jaskot	& LCVV05010 & G130M	& 1327 & 8096  \\[1.5ex]	
		        & 16643 & Gazagnes	& LEQR05010 & G160M	& 1533 & 7551  \\	
\tableline
17. J1712$+$3216& 14628 & Berg	    & LDAF06010 & G140L	& 1280 & 2076  \\[1.5ex]	
		        & 16677 & Stark	    & LENE15010 & G160M	& 1533 & 7678 \\	
				&       &           & LENE16010	&       &      & 7678  \\	
\tableline
18. J2238$+$1400& 16643 & Gazagnes	& LEQR01010 & G130M	& 1291 & 7505  \\
				&       &           & LEQR02020	&       &      & 2691  \\[1.5ex]	
		        & 16677 & Stark	    & LENE17010 & G160M	& 1533 & 7607  \\	
				&       &           & LENE55010	&       &      & 4914  \\
\enddata
\tablecomments{
Details of the archival and new \textit{HST}/COS observations of our \ion{C}{4} emission sample are listed in this table. Columns 2 and 3 give the program IDs (PIDs) and principal investigators (PIs), respectively. For each program, Columns 4--7 list the dataset ID, grating, central wavelength, and exposure time. Exposure times are taken from the STScI MAST observation catalog (\texttt{t\_exptime}) for each listed dataset and rounded to the nearest second. The table includes FUV (G130M, G160M, and G140L) and NUV (G185M) observations. The velocity-resolved \ion{C}{4} fits use the medium-resolution FUV spectra; the G140L and G185M observations are listed as ancillary data.}
\end{deluxetable*}

\section{Inferred Radiative Transfer Parameters}
\label{sec:rt_parameters}

Table~\ref{tbl:civ_rt_parameters} lists the inferred
\ion{C}{4} RT parameters for all 18 galaxies. We report
posterior medians and uncertainties spanning the
16th--84th percentiles. Derived quantities, including the
LOS \ion{C}{4} column density and maximum outflow velocity,
are calculated for each posterior sample before computing
weighted percentiles, thereby retaining the correlations
among the underlying model parameters. 

\begin{deluxetable*}{lcccccccccc}
\tabletypesize{\scriptsize}
\tablewidth{0pt}
\setlength{\tabcolsep}{2pt}
\tablecaption{Inferred \ion{C}{4} Radiative Transfer Parameters\label{tbl:civ_rt_parameters}}
\tablehead{
\colhead{Target} & \colhead{$\log N_{\rm CIV,LOS}$} &
\colhead{$b_{\rm D,cl}$} & \colhead{$\sigma_{\rm cl}$} &
\colhead{$v_0$} & \colhead{$R$} & \colhead{$v_{\rm out,max}$} &
\colhead{$\mathrm{EW}_{\rm int}$} & \colhead{$\tau_{\rm d,cl}$} &
\colhead{$b_{\max}/R_{\rm out}$} & \colhead{$\Delta v$} \\[-1ex]
\colhead{} & \colhead{($\mathrm{cm}^{-2}$)} &
\colhead{($\mathrm{km\,s^{-1}}$)} & \colhead{($\mathrm{km\,s^{-1}}$)} &
\colhead{($\mathrm{km\,s^{-1}}$)} & \colhead{} &
\colhead{($\mathrm{km\,s^{-1}}$)} & \colhead{(\AA)} &
\colhead{($10^{-3}$)} & \colhead{} & \colhead{($\mathrm{km\,s^{-1}}$)}
}
\startdata
J0337$-$0502 & $16.59^{+0.09}_{-0.08}$ & $10.3^{+0.4}_{-0.4}$ & $56.8^{+2.9}_{-3.7}$ & $74.8^{+5.2}_{-6.4}$ & $1.54^{+0.02}_{-0.02}$ & $24.7^{+2.0}_{-2.3}$ & $4.79^{+0.40}_{-0.53}$ & $6.17^{+0.80}_{-1.13}$ & $0.319^{+0.069}_{-0.042}$ & $11.2^{+1.5}_{-1.5}$ \\
J0825$+$3532 & $16.50^{+0.33}_{-0.33}$ & $14.1^{+1.7}_{-1.7}$ & $35.0^{+3.0}_{-2.8}$ & $28.4^{+6.6}_{-4.7}$ & $2.19^{+0.10}_{-0.13}$ & $18.1^{+2.6}_{-2.3}$ & $3.32^{+0.50}_{-0.50}$ & $4.86^{+1.71}_{-2.78}$ & $0.783^{+0.195}_{-0.221}$ & $-6.6^{+3.1}_{-3.0}$ \\
J0842$+$1033 & $16.55^{+0.40}_{-0.44}$ & $11.8^{+2.8}_{-2.2}$ & $25.2^{+5.1}_{-5.2}$ & $35.2^{+13.6}_{-10.0}$ & $2.07^{+0.19}_{-0.12}$ & $21.0^{+6.3}_{-5.3}$ & $6.28^{+0.82}_{-0.76}$ & $4.23^{+2.44}_{-2.80}$ & $0.747^{+0.172}_{-0.195}$ & $12.2^{+4.4}_{-4.6}$ \\
J0934$+$5514 & $17.46^{+0.37}_{-0.38}$ & $17.9^{+4.6}_{-1.6}$ & $72.4^{+5.1}_{-9.9}$ & $49.6^{+8.0}_{-7.6}$ & $2.83^{+0.32}_{-0.20}$ & $44.4^{+4.6}_{-4.6}$ & $1.41^{+0.39}_{-0.28}$ & $0.96^{+6.12}_{-0.76}$ & $0.290^{+0.035}_{-0.039}$ & $-19.3^{+4.1}_{-3.7}$ \\
J0947$+$4138 & $15.38^{+0.75}_{-0.47}$ & $23.5^{+13.5}_{-11.4}$ & $31.3^{+7.7}_{-6.1}$ & $46.4^{+12.9}_{-12.8}$ & $2.44^{+0.24}_{-0.24}$ & $33.7^{+7.0}_{-7.6}$ & $2.07^{+0.50}_{-0.40}$ & $2.60^{+1.75}_{-1.78}$ & $0.683^{+0.253}_{-0.236}$ & $12.6^{+9.7}_{-28.7}$ \\
J0954$+$0952 & $16.64^{+0.40}_{-0.40}$ & $9.0^{+1.9}_{-1.7}$ & $76.8^{+19.0}_{-20.5}$ & $22.6^{+17.6}_{-15.2}$ & $2.17^{+0.31}_{-0.28}$ & $14.2^{+10.0}_{-9.7}$ & $4.98^{+0.77}_{-0.69}$ & $2.63^{+1.98}_{-1.83}$ & $0.722^{+0.241}_{-0.254}$ & $-1.8^{+5.7}_{-6.8}$ \\
J1044$+$0353 & $16.62^{+0.09}_{-0.09}$ & $17.8^{+0.7}_{-0.6}$ & $52.0^{+1.3}_{-1.3}$ & $38.8^{+2.0}_{-1.8}$ & $2.47^{+0.04}_{-0.04}$ & $29.3^{+1.1}_{-1.0}$ & $10.40^{+0.30}_{-0.27}$ & $0.49^{+0.11}_{-0.09}$ & $0.802^{+0.114}_{-0.113}$ & $-3.6^{+1.2}_{-1.3}$ \\
J1131$+$5703 & $18.05^{+0.29}_{-0.43}$ & $7.6^{+2.4}_{-1.6}$ & $32.9^{+5.9}_{-4.7}$ & $32.2^{+12.7}_{-7.9}$ & $2.38^{+0.28}_{-0.26}$ & $23.2^{+5.4}_{-4.3}$ & $4.13^{+0.99}_{-0.63}$ & $4.84^{+2.24}_{-3.22}$ & $0.569^{+0.279}_{-0.255}$ & $-10.6^{+6.6}_{-7.1}$ \\
J1201$+$0211 & $16.56^{+0.39}_{-0.40}$ & $12.1^{+3.7}_{-3.4}$ & $36.0^{+4.2}_{-4.3}$ & $34.6^{+8.4}_{-6.6}$ & $2.13^{+0.13}_{-0.12}$ & $21.1^{+3.8}_{-3.2}$ & $2.38^{+0.60}_{-0.37}$ & $2.34^{+1.94}_{-1.61}$ & $0.354^{+0.075}_{-0.086}$ & $-16.2^{+7.8}_{-5.9}$ \\
J1202$+$5416 & $15.71^{+0.48}_{-0.35}$ & $13.5^{+3.4}_{-3.1}$ & $61.6^{+6.4}_{-6.0}$ & $69.4^{+13.7}_{-14.0}$ & $2.34^{+0.30}_{-0.19}$ & $48.7^{+6.4}_{-6.5}$ & $1.92^{+0.50}_{-0.32}$ & $2.83^{+1.85}_{-1.96}$ & $0.319^{+0.060}_{-0.063}$ & $16.0^{+6.4}_{-8.9}$ \\
J1214$+$5345 & $15.80^{+0.59}_{-0.41}$ & $6.8^{+2.0}_{-1.4}$ & $38.9^{+4.0}_{-3.6}$ & $61.6^{+12.7}_{-13.7}$ & $2.26^{+0.24}_{-0.14}$ & $40.9^{+5.4}_{-6.1}$ & $3.36^{+0.34}_{-0.27}$ & $2.13^{+1.28}_{-1.55}$ & $0.712^{+0.189}_{-0.210}$ & $22.2^{+1.9}_{-3.2}$ \\
J1323$-$0132 & $16.47^{+0.06}_{-0.06}$ & $18.5^{+1.1}_{-1.2}$ & $49.7^{+1.5}_{-1.6}$ & $46.7^{+4.5}_{-3.6}$ & $2.20^{+0.06}_{-0.09}$ & $29.9^{+1.6}_{-1.6}$ & $9.93^{+0.83}_{-0.67}$ & $2.37^{+0.43}_{-0.45}$ & $0.329^{+0.048}_{-0.043}$ & $-12.3^{+2.0}_{-2.0}$ \\
J1325$-$1136 & $17.37^{+0.37}_{-0.47}$ & $19.3^{+2.9}_{-2.6}$ & $68.0^{+4.3}_{-4.4}$ & $34.7^{+13.2}_{-11.0}$ & $2.15^{+0.42}_{-0.18}$ & $22.1^{+4.7}_{-4.4}$ & $7.00^{+0.39}_{-0.38}$ & $0.73^{+0.90}_{-0.52}$ & $0.769^{+0.156}_{-0.188}$ & $-19.2^{+3.1}_{-2.9}$ \\
J1331$+$4151 & $15.80^{+0.38}_{-0.26}$ & $11.8^{+2.3}_{-2.1}$ & $52.5^{+4.0}_{-3.8}$ & $40.5^{+9.4}_{-6.6}$ & $2.73^{+0.30}_{-0.36}$ & $34.4^{+4.1}_{-3.6}$ & $1.24^{+0.31}_{-0.18}$ & $2.38^{+0.48}_{-1.66}$ & $0.426^{+0.104}_{-0.098}$ & $23.2^{+1.3}_{-2.6}$ \\
J1418$+$2102 & $16.28^{+0.37}_{-0.40}$ & $21.4^{+4.8}_{-4.2}$ & $159.3^{+32.5}_{-35.8}$ & $38.9^{+20.0}_{-23.9}$ & $2.12^{+0.25}_{-0.26}$ & $22.7^{+12.0}_{-14.0}$ & $2.35^{+0.77}_{-0.55}$ & $1.50^{+1.01}_{-0.99}$ & $0.734^{+0.179}_{-0.193}$ & $-1.7^{+9.3}_{-9.7}$ \\
J1509$+$3731 & $18.85^{+0.16}_{-0.19}$ & $13.0^{+4.5}_{-4.1}$ & $48.3^{+8.5}_{-9.0}$ & $48.9^{+18.0}_{-13.4}$ & $2.24^{+0.44}_{-0.27}$ & $32.1^{+5.5}_{-4.6}$ & $2.81^{+0.55}_{-0.45}$ & $1.61^{+0.99}_{-1.17}$ & $0.448^{+0.100}_{-0.116}$ & $-13.7^{+7.9}_{-6.6}$ \\
J1712$+$3216 & $18.35^{+0.26}_{-0.49}$ & $11.7^{+3.3}_{-2.4}$ & $203.9^{+27.4}_{-24.6}$ & $55.8^{+27.1}_{-21.7}$ & $2.01^{+0.27}_{-0.21}$ & $31.5^{+11.9}_{-11.9}$ & $4.42^{+1.81}_{-1.19}$ & $1.95^{+1.36}_{-1.31}$ & $0.533^{+0.289}_{-0.187}$ & $21.4^{+2.5}_{-3.7}$ \\
J2238$+$1400 & $16.38^{+0.63}_{-0.67}$ & $9.1^{+3.0}_{-2.6}$ & $35.4^{+3.7}_{-3.7}$ & $63.6^{+15.2}_{-12.3}$ & $2.38^{+0.18}_{-0.19}$ & $45.4^{+6.7}_{-6.8}$ & $4.73^{+0.54}_{-0.45}$ & $2.80^{+1.71}_{-1.89}$ & $0.713^{+0.237}_{-0.231}$ & $19.2^{+4.2}_{-6.8}$ \\
\enddata
\tablecomments{Values are posterior medians with uncertainties spanning the
16th--84th percentiles (approximately $1\sigma$). Derived quantities are
evaluated for each posterior sample before computing weighted percentiles.
$N_{\rm CIV,\,LOS}$ is the line-of-sight C$^{3+}$ column density;
$b_{\rm D,\,cl}$ is the clump Doppler parameter; $\sigma_{\rm cl}$ is the clump
velocity dispersion; $v_0$ and $R$ parameterize the adopted outflow velocity law,
with $v_{\rm out,\,max}=v_0\sqrt{R-1-\ln R}$ \citep{Li2026a}.
$\mathrm{EW}_{\rm int}$ is the intrinsic rest-frame \ion{C}{4} doublet equivalent width, defined as positive for emission and calculated using an injected continuum bandwidth of $1500\,\mathrm{km\,s^{-1}}$;
$\tau_{\rm d,\,cl}$ is the
clump dust optical depth; $b_{\max}/R_{\rm out}$ is the effective aperture radius relative to the outer model radius; and $\Delta v$ is the fitted velocity offset. The dust optical depths are tabulated in units of $10^{-3}$, and aperture ratios are expressed as fractions rather than percentages.}
\end{deluxetable*}

\facilities{HST (COS)}

\software{CalCOS (STScI; \url{https://github.com/spacetelescope/calcos}),
          Starburst99 \citep{leitherer99},
          Cloudy \citep{ferland13,ferland17},
          MPFIT \citep{markwardt09},
          PEACOCK \citep{Li2026a},
          TensorFlow \citep{Abadi2016},
          Keras \citep{Chollet2015},
          dynesty \citep{Speagle2020}}

\end{document}